\documentclass[12pt]{article}

\def\thefootnote{\fnsymbol{footnote}}
\def\bea{\begin{eqnarray}}
\def\eea{\end{eqnarray}}
\def\beq{\begin{equation}}
\def\eeq{\end{equation}}

\def\bL{\bar{\Lambda}}
\def\ux{$U(1)_X$}
\def\vx{V_X}

\def\const{{1\over32\pi^2}}
\def\superint{\int d^{4}\theta}

\newcommand{\DaDa}{{\cal D}^{\alpha}{\cal D}_{\alpha}}
\newcommand{\DbDb}{{\cal D}_{\dot{\alpha}}{\cal D}^{\dot{\alpha}}}
\newcommand{\Da}{{\cal D}_{\alpha}}
\newcommand{\Db}{{\cal D}^{\dot{\beta}}}
\newcommand{\Dc}{{\cal D}^{\alpha}}
\newcommand{\Dd}{{\cal D}_{\dot{\beta}}}
\newcommand{\Wa}{W_{\alpha}}
\newcommand{\Wb}{W^{\dot{\beta}}}
\newcommand{\Wc}{W^{\alpha}}
\newcommand{\Wd}{W_{\dot{\beta}}}
\newcommand{\Xa}{X_{\alpha}}
\newcommand{\ka}{k_{\alpha}}

\newcommand{\kc}{k^{\alpha}}

\def\dotb{\dot{\beta}}

\def\hL{\hat{L}}

\def\re{{\rm Re}}
\def\im{{\rm Im}}
\def\btr{{\bf Tr}}
\def\bM{\bar{M}}
\def\bN{\bar{N}}
\def\bC{\bar{C}}
\def\bB{\bar{B}}

\def\W{\overline{W}}
\def\btA{\bar{\tilde A}}
\def\G{{\cal G}}
\def\cW{{\cal W}}
\def\cbW{{\cal\W}}
\def\tD{{\tilde D}}

\def\tr{{\tilde r}}
\def\tG{{\tilde G}}

\def\tGa{{\tilde\Gamma}}

\def\tF{{\tilde F}}
\def\tZ{{\widetilde Z}}
\def\tY{{\widetilde Y}}
\def\tR{{\tilde R}}
\def\tL{{\tilde L}}
\def\tA{{\tilde A}}
\def\tK{{\tilde K}}
\def\tN{{\tilde N}}
\def\tL{{\tilde L}}

\def\hG{{\hat G}}
\def\hZ{{\widehat Z}}
\def\hY{{\widehat Y}}
\def\hGa{{\hat\Gamma}}

\def\hR{{\hat R}}
\def\hL{{\hat L}}
\def\hK{{\hat K}}
\def\D{{\cal D}}

\def\bD{\bar{\D}}

\def\pp{\partial}

\def\lln{{\ln\Lambda^2\over 32\pi^2}}
\def\ibar{\bar{\imath}}

\def\bbet{\bar{\beta}}
\def\[{\left [}
\def\]{\right ]}
\def\({\left (}
\def\){\right )}
\def\lbr{\left\{}
\def\rbr{\right\}}
\def\r{\right|}
\def\l{\left.}

\def\J{\bar{J}}
\def\H{\bar{H}}

\def\T{\bar{T}}
\def\p{\bar{p}}

\def\z{\bar{z}}

\def\R{{\cal{R}}}
\def\S{{\bar{S}}}
\def\Det{{\rm Det}}
\def\STr{{\rm STr}}
\def\Tr{{\rm Tr}}

\def\K{{\cal K}}
\def\Gi{G^{(i)}}
\newcommand{\Gia}{\Gi_{\alpha}}
\newcommand{\Gic}{G_{(i)}^{\alpha}}

\newcommand{\Gaa}{\Gamma_{\alpha}}

\def\Ti^{T^{(i)}}
\def\Gj{G^{(j)}}
\def\Gii{G_{(i)}}

\def\Fi{F^{(i)}}
\def\bFi{\bF^{(i)}}

\def\f{\bar{f}}
\def\bR{\bar{R}}

\def\L{{\cal L}}
\def\R{{\cal{R}}}

\def\hf{\hat{\phi}}

\def\hA{\hat{A}}

\def\hG{\hat{G}}
\def\hV{\hat{V}}
\def\n{\bar{n}}
\def\m{\bar{m}}
\def\s{\bar{s}}

\def\bth{\bar{\theta}}

\def\tM{{\widetilde M}}

\def\A{\bar{A}}

\def\Y{{\bar{Y}}}
\def\Z{{\bar{Z}}}

\def\bv{\bar{\varphi}}
\def\bph{\bar{\phi}}
\def\bPh{\bar{\Phi}}
\def\hph{\hat{\varphi}}
\def\hbp{\hat{\bv}}

\def\tph{\tilde{\varphi}}
\def\tbp{\tilde{\bv}}

\def\bF{\bar{F}}

\def\bj{\bar{\jmath}}
\def\a0{\alpha_0}

\begin{document}

\begin{titlepage}
\begin{center}

\hfill LBNL-44372 \\
\hfill UCB-PTH-99/47 \\
\hfill hep-th/9910147 \\
\hfill October 1999 \\[.3in]

{\large {\bf ONE-LOOP REGULARIZATION OF SUPERGRAVITY II: THE DILATON
AND THE SUPERFIELD FORMULATION}}\footnote{This work was supported in
part by the Director, Office of Science, Office of High Energy and
Nuclear Physics, Division of High Energy Physics, of the
U.S.Department of Energy under Contract DE-AC03-76SF00098 and in part
by the National Science Foundation under grant
PHY-95-14797.}\footnote{This paper is dedicated to the memory of
Kamran Saririan.} \\[.2in]

Mary K. Gaillard \\[.1in]

{\em Department of Physics and Theoretical Physics Group,
 Lawrence Berkeley Laboratory, 
 University of California, Berkeley, California 94720}\\[.5in] 

\end{center}

\begin{abstract}

The on-shell regularization of the one-loop divergences of
supergravity theories is generalized to include a dilaton of the type
occurring in effective field theories derived from superstring theory,
and the superfield structure of the one-loop corrections is given.
Field theory anomalies and quantum contributions to soft supersymmetry
breaking are discussed.  The latter are sensitive to the precise
choice of couplings that generate Pauli-Villars masses, which in turn
reflect the details of the underlying theory above the scale of the
effective cut-off.  With a view to the implementation the Green-Schwarz
and other mechanisms for canceling field theory anomalies under a
$U(1)$ gauge transformation and under the T-duality group of modular
transformations, we show that the K\"ahler potential renormalization
for the untwisted sector of orbifold compactification can be made
invariant under these groups.

\end{abstract}
\end{titlepage}

\newpage

\renewcommand{\thepage}{\roman{page}}
\setcounter{page}{2}
\mbox{ }

\vskip 1in

\begin{center}
{\bf Disclaimer}
\end{center}

\vskip .2in

\begin{scriptsize}
\begin{quotation}
This document was prepared as an account of work sponsored by the United
States Government. While this document is believed to contain correct 
 information, neither the United States Government nor any agency
thereof, nor The Regents of the University of California, nor any of their
employees, makes any warranty, express or implied, or assumes any legal
liability or responsibility for the accuracy, completeness, or usefulness
of any information, apparatus, product, or process disclosed, or represents
that its use would not infringe privately owned rights.  Reference herein
to any specific commercial products process, or service by its trade name,
trademark, manufacturer, or otherwise, does not necessarily constitute or
imply its endorsement, recommendation, or favoring by the United States
Government or any agency thereof, or The Regents of the University of
California.  The views and opinions of authors expressed herein do not
necessarily state or reflect those of the United States Government or any
agency thereof, or The Regents of the University of California.
\end{quotation}
\end{scriptsize}

\vskip 2in

\begin{center}
\begin{small}
{\it Lawrence Berkeley Laboratory is an equal opportunity employer.}
\end{small}
\end{center}

\newpage
\renewcommand{\theequation}{\arabic{section}.\arabic{equation}}
\renewcommand{\thepage}{\arabic{page}}
\setcounter{page}{1}
\def\thefootnote{\arabic{footnote}}
\setcounter{footnote}{0}

\section{Introduction}
\hspace{0.8cm}\setcounter{equation}{0} 
It has been
shown~\cite{pv1}--\cite{pvcan}, that Pauli-Villars (PV) regularization
of one-loop ultraviolet divergences is possible for an $N=1$
supergravity theory if Yang-Mills fields have canonical kinetic
energy.  In this paper those results are generalized to include their
couplings to a dilaton.  In Section 2 we summarize earlier results,
and display the logarithmically divergent one-loop corrections in the
form of superfield operators, which permits the extension of those
results to fermionic terms~\cite{dela,bpz} in the one-loop corrected
effective Lagrangian.  This formulation will also be convenient for
the subsequent analysis.  In Section 3 the dilaton is incorporated in
the Pauli-Villars regularization of anomaly-free supergravity
described in Ref.~\cite{pvcan}, hereafter referred to as I. The
application of PV regularization to determine soft supersymmetry
breaking terms is also discussed in this section. It is shown that the
contributions to A-terms are highly sensitive to the details of the
regularization.  In Section 4 we
regulate effective theories of orbifold compactification with twisted
sector fields set to zero in the background.  We show that this
regularization can be done in such a way that the renormalization of
the K\"ahler potential is invariant under modular (T-duality)
transformations; we have in mind the construction of an effective
one-loop Lagrangian that is perturbatively modular invariant.  In
Section 5 the discussion of regularization and anomalies is extended
to theories with an anomalous $U(1)$ gauge symmetry. The results are
summarized in Section 6, where we discuss issues still to be
addressed in order to achieve full anomaly cancellation.  Many
calculational details are relegated to the appendixes. 

\section{Preliminaries}
\hspace{0.8cm}\setcounter{equation}{0}

In this paper we consider supergravity theories defined by the
standard Lagrangian~\cite{crem,bggm} with $N$ chiral multiplets $Z^i =
\Phi^1,...\Phi^{N-1},S$, where $S$ is a gauge singlet, and $N_G$ gauge
supermultiplets.  The K\"ahler potential $K$, superpotential $W$
and gauge kinetic function $f$ are given by
\bea K(Z,\Z) &=& - \ln(S +\S) + G(\Phi,\bPh) = k + G, \quad W(Z) = 
W(\Phi), \nonumber \\f_{ab}(Z) &=& \delta_{ab}S = \delta_{ab}(x+iy), 
\label{sdil}\eea
which are the classical functions found in string compactifications with
affine level one.\footnote{The results can be generalized to the 
case $f_{ab} = \delta_{ab}k_af,\;k_a=$ constant, 
by making the substitutions $F^a_{\mu\nu}\to k_a^{1\over2}F^a_{\mu\nu}, 
\;A^a_\mu\to k_a^{1\over2}A^a_\mu, \; T^a\to k_a^{-{1\over2}}T^a.$}
In this section we briefly recall the results of~\cite{pv1,pvcan}, and 
cast them in a superfield form that will allow us to short-cut some 
of the subsequent calculations.

\subsection{One-loop logarithmic divergences in supergravity}
The ultra-violet divergent part of the one-loop corrected supergravity
Lagrangian for bosons was calculated in~\cite{us}-\cite{us2}.  The result for 
the logarithmically divergent contribution is
\bea \L_{eff} &=& \L\(g_R,K_R\) + \sqrt{g}\lln L \nonumber \\ L &=&
\tL_0 + L'_0 + \tL_1 + L_2 + L_3 + NL_\chi + N_G(\tL_g + L'_g),
\nonumber \\ \tL_0 &=& L_0 + 41L_{GB}, \quad \tL_\chi = 
L_\chi + L_{GB}, \quad \tL_g = L_g - 3L_{GB}, \nonumber \\
K_R &=& K + \lln\[e^{-K}A_{ij}\A^{ij} -2\hV + (N_G - 10)M^2 - 4\K^a_a 
-16\D\],\nonumber \\ \K^a_b &=& {1\over x}(T^az)^i(T_b\z)^{\m}K_{i\m},
\quad A = e^KW = \A^{\dag}, \;\;\;\; A_{ij} = D_iD_jA. \label{li}\eea
where $\L(g,K)$ is the standard  
Lagrangian~\cite{crem,bggm} for $N=1$ supergravity coupled to
matter with space-time metric $g_{\mu\nu}$, K\"ahler potential $K$
and superpotential $W$.  
$V = \hV + \D$ is the classical scalar potential with $\hV = e^{-K}A_i\A^i - 3M^2,$
$A_i = D_iA,$ $\D = (2x)^{-1}\D^a\D_a,$ $\D_a = K_i(T_az)^i$,
$M^2 = e^{-K}A\A$ is the field-dependent squared gravitino mass, and
$D_i$ is the scalar field reparameterization covariant derivative.
Scalar indices are lowered and raised with
the K\"ahler metric $K_{i\m}$ and its inverse $K^{i\m}$.

The operators $L_A$ in (\ref{li}) are given in component 
form\footnote{See Appendix D of I and Appendix E below 
for corrections to~\cite{us,us2}. There is an
extraneous factor of $x$ in the second line of (2.26) in I.} in
Eqs. (2.25--27) of I, 
\beq L_{GB} = {1\over48} \(r^{\mu\nu\rho\sigma}r_{\mu\nu\rho\sigma} - 
4r^{\mu\nu}r_{\mu\nu} + r^2\), \eeq
is the Gauss-Bonnet term which is a total derivative, and was not included
explicitly in I.  The operators $L'_A$ are additional contributions that
arise in the presence of a dilaton coupling to the Yang-Mills terms.
Their component field expressions read: 
\bea L'_0 &=& 92\D M^2 
- 2x^2\cW_{ab}\cbW^{ab} - 4x^2\cW\cbW \nonumber \\ & & 
- {\pp_\rho s\pp^\nu\s\over x}F^{+a}_{\mu\nu}
F_{-a}^{\mu\rho} + 10{\pp_\mu s\pp^\mu\s\over x^2}\D + 
4i{\pp_\mu s\pp_\nu\s\over x^2}\D^aF_a^{\mu\nu}
\nonumber \\ & & - {6\over x}\lbr\[i\pp_\nu sF_{-a}^{\nu\mu} 
+ {\pp^\mu s\over x}\D_a\]\D_\mu\z^{\m}K_{i\m}(T^az)^i + 
{\rm h.c.}\rbr \nonumber \\ & &
+ xF^{-a}_{\rho\mu}F_{+a}^{\rho\nu}\D_\nu z^i\D^\mu\z^{\m}K_{i\m} 
+ 2i\D_\mu z^i\D_\nu\z^{\m}K_{i\m}\D^aF_a^{\mu\nu} \nonumber \\ & & 
+ 4\D\hV + 2\D K_{i\m}\D_\rho z^i\D^\rho\z^{\m}  ,\\
L'_g &=& - x\(\cW + \cbW\)\(M^2 + \hV\) 
- {2\over3}M^2\(\D_\mu z^i\D^\mu\z^{\m}K_{i\m} + 4\hV - 2\D\)
\nonumber \\ & & - 7M^4 + {\pp_\mu s\pp^\mu s\pp_\nu\s\pp^\nu\s \over16x^4}
- {\pp_\mu s\pp_\nu\s\over2x^2}K_{i\m}\(\D^\mu z^i
\D^\nu\z^{\m} + \D^\mu\z^{\m}\D^\nu z^i\) \nonumber \\ & & 
+ x^2\cW\cbW + \[F^{+a}_{\rho\mu}F_{-a}^{\rho\nu}
+ {2\over3}g^\nu_\mu\(2K_{i\m}\D_\rho z^i\D^\rho\z^{\m} - \hV - \D\)\]
{\pp_\nu s\pp^\mu\s\over4x}\nonumber \\ & &
+ {e^{-K}\over2x}\(\pp_\mu\s\D^\mu z^iA_i\A + {\rm h.c.}\), \label{pvls}\eea 
where 
\bea \cW_{ab} &=& {1\over4}\(F_a\cdot F_b - i\tF_a\cdot F_b\)
- {1\over2x}\D_a\D_b = - {1\over2}\l\DaDa W^\beta W_\beta\r,
\nonumber \\ F_{a\nu\mu}^{\mp} &=& F_{a\nu\mu}\mp i\tF_{a\nu\mu},
\quad x = {\rm Re}s, \quad \cW = \cW^a_a, \eea
with $F^a_{\mu\nu}$ the Yang-Mills field strength.  As in I we have
dropped total derivatives (except for the Gauss-Bonnet term)
and other terms that do not contribute to the
S-matrix, by virtue of the classical equations of motion of the
physical fields.

  It will be convenient here to display these operators in
superfield form. $\theta$-integration of the superfield operators
gives expressions that include the various auxiliary fields.
Replacing these by the solutions of their classical equations of
motion gives the component expressions, up to terms that do not
contribute to the S-matrix. We will display here the component
expressions only for those operators that are not included in I.
The component expressions for operators constructed from tensor-valued
functions $T(Z,\Z)$ are given in Appendix A.

In the K\"ahler $U(1)$ superspace formulation of
supergravity, a general ``F-term'' Lagrangian takes the form~\cite{bggm}
\beq L_A = L(\Phi_A) = {1\over2}\superint{E\over R}\Phi_A + 
{\rm h.c.},\label{fterm} \eeq
where $\Phi$ is a chiral superfield of K\"ahler $U(1)$ weight $w(\Phi) = 2$.
Here we construct these fields as bilinears in 
chiral superfields of weight 1, namely 
the Yang-Mills field strength superfield $W^a_{\alpha}$,
the curvature superfield $W_{\alpha\beta\gamma}$ (the lowest
components of the totally symmetrized spinorial derivatives $\l
\D_{\{\gamma}W_{\alpha\beta\gamma\}}\r$ are elements of the Riemann 
tensor), and the superfields
\beq T_\alpha = - {1\over8}\(\DbDb - 8R\)\hat{T}_\alpha, \quad
\hat{T}_\alpha = T_i\D_\alpha Z^i,
\label{tft} \eeq
where $T_i(Z,\Z)$ is any (tensor-valued) zero-weight
function of the chiral and
anti-chiral superfields.  In particular, the chiral superfield 
\beq K_\alpha = X_\alpha = - {1\over8}\(\DbDb - 8R\)\D_\alpha K,
\label{defx} \eeq
was introduced in~\cite{bggm}; the lowest 
component of its spinorial derivative $-{1\over2}\D^\alpha X_\alpha |$ 
is the kinetic term for matter fields in the classical Lagrangian.
Then defining
\beq \Phi_W  =  {1\over6}W^{\alpha\beta\gamma}W_{\alpha\beta\gamma}, \quad
 \Phi_{YM}^a = {1\over4}W_a^\alpha W^a_\alpha, \quad 
 \Phi_\alpha = - {1\over2} X^\beta X_\beta, \label{ops}\eeq
we may write (see Appendix A), up to total derivatives and field redefinitions,
\bea \tL_0 &=& 41\tL_\chi + 6\(L_\chi - C_aL^a_{YM} + \hL_0\) -
{20\over3}L_\alpha, \nonumber \\ \tL_\chi &=& L_W + {1\over2}L_\chi + {1\over9}L_\alpha,
\quad \tL_G = - 3\tL_\chi + 6L_\chi - {1\over3}L_\alpha, \label{lops} \eea
where $C_a$ is the quadratic Casimir in the adjoint representation
of the gauge subgroup $\G_a:\;\Tr(T_aT_b)_{\rm adj} = \delta_{ab}C_a$ 
with $T_a$ a generator of $\G_a$ and $T_b$ any generator. $L_\alpha$ is 
given in component form in (2.40) of I. 
The operators $L_\chi$ and
\bea \hL_0 &=& \(\hV + 2M^2\) K_{i\m} \D_\mu\z^{\m}\D^\mu z^i 
+   M^2\(2\hV + 3M^2  + 2\D\) \nonumber \\ & &
+ \D_\mu z^j\D^\mu z^i\D_\nu\z^{\m}\D^\nu\z^{\n}K_{i\n}K_{j\m}\eea
are ``D-terms'' of the form
\beq L_A = L(\phi_A) = \superint E\phi_A = - {1\over16}\superint{E\over R}\(\bD^2 -
8R\)\phi_A + {\rm h.c.}, \quad w{(\phi_A)=0}. \label{dterm}\eeq
To include these we define the zero-weight real superfields
\bea T^{\alpha\dot{\beta}}_{\alpha\dot{\beta}} &=& {1\over16}\Dc Z^i\Da Z^j
\Dd\Z^{\m}\Db\Z^{\n}T_{ij\m\n},\nonumber \\ 
\phi_{WT} &=& {x\over2}\Wc_a\Da Z^i\Wd^a\Db\Z^{\m}T_{i\m}, \quad
T^\alpha_\alpha = {1\over2}\Dc Z^i\Da Z^jT_{ij} + {\rm h.c.},
\nonumber \\ \phi_{\cW^a_b} &=& {x^2\over4}\Wc_a\Wa^b\Wd^a\Wb_b, \quad
\phi_{\cW} = {x^2\over4}\Wc_a\Wa^a\Wd^b\Wb_b. \label{tdt}\eea
With these definitions we have
\bea \phi_\chi &=& {1\over3}\hf_0 - {1\over6}\phi_{WK}
+ {1\over3}\phi_{\cW^a_b},\quad
\hf_0 = K^{\alpha\dot{\beta}}K_{\alpha\dot{\beta}} - 
e^K|W(Z)|^2. \eea
The last term in $\hf_0$ is equivalent to a renormalization of the 
K\"ahler potential; up to a field-dependent Weyl scaling and higher
order terms in the loop expansion parameter,
the shift in $\L/\sqrt{g}$ due to a shift $F(Z,\Z)$ in the
K\"ahler potential is given by
\bea {1\over\sqrt{g}}\Delta_F\L &=& \Delta_F L = - F\hV + 
\(e^{-K}\A^iA^{\m} + \D_\mu z^i\D^\mu\z^{\m}\)\pp_i\pp_{\m}F 
\nonumber \\ & &
- \lbr \pp_iF\[e^{-K}\A^i A + {1\over2x}\D_a(T^az)^i\] + {\rm h.c.}\rbr
= {1\over\sqrt{g}}\int d^4\theta E F.\label{delK}\eea
As shown in Appendix A, $L_\chi$ can be obtained as a linear
combination of $L_\alpha$ and an operator generated by a metric field
redefinition that eliminates terms quadratic in the space-time scalar
curvature and the Ricci tensor.  That is, it is equivalent to a linear
combination of $L_\alpha$ and a D-term (\ref{dterm}) constructed from
the superfields that determine the elements of the super-Riemann and
torsion tensors~\cite{bggm}: $\phi_a = R\bR,\; G_aG^a,\ldots$. In
addition we have the F-terms $L_1,L_2$ with 
\bea \Phi_1 &=& 2C_a^M\Phi^a_{YM} - {1\over2}
\Gamma_j^{i\alpha}\[\Gamma^j_{i\alpha} + 2(T_a)^j_iW^a_\alpha\] 
, \nonumber \\
\Phi_2 &=& {1\over3}X^\alpha\[\Gamma_\alpha + 2(T_a)^i_iW^a_\alpha 
\], \quad 
\Gamma_\alpha = \Gamma^i_{i\alpha}, \label{rho}\eea
where $Z^i$ is a matter chiral superfield ($w(Z) = 0$),
$\Gamma^i_{jk}$ is an element of the affine connection associated with
the K\"ahler metric, and $C^M_a$ is the matter quadratic Casimir for
the gauge subgroup $\G_a$: $(T_aT_b)^i_i = \delta_{ab}C^M_a.$ These
contributions to (\ref{li}) are canceled by identical contributions
from negative signature PV chiral superfields $Z^I$ with the same
gauge charges and K\"ahler metric as the matter fields.

The terms proportional to
$L_{0,\chi,g}$ are partially canceled by the introduction of PV chiral 
superfields $\phi^C$ with K\"ahler metric 
\beq K_{C\bC} = e^{\alpha_CK}, \quad \Gamma^C_{Di} =
\alpha_C\delta^C_DK_i, \quad \Gamma^C_{D\alpha} =
\alpha_C\delta^C_DX_\alpha,\label{rhoc}\eeq
some of which carry gauge charge.  Assuming $\sum_C(T_a)^C_C = 0$, the
$\phi^C$-loop gives a contribution: \beq \(L_1 + L_2\)_{\phi^C} =
\eta^C\[2C_aL_{YM}^a + \( \alpha^C - {2\over3}\)\alpha^C
L_\alpha\], \label{lalph}\eeq   
where $\eta^C = \pm 1$ denotes the signature of the PV field
$\Phi^C$. The operator $L_3$ depends both on elements $R_{i\m j\n}$ 
of the K\"ahler Riemann tensor and on covariant scalar derivatives of
$A = e^KW$; it is the bosonic part of a 
D-term\footnote{Note that $T_{kl} = e^{-K/2}A_{kl}(Z,\Z)$ is a
superfield of weight $w(T_{kl}) = 2$; its spinorial derivatives
satisfy $\Db T_{kl} = e^{K/2}\Db\Z^{\m}D_{\m} \(e^{-K}A_{kl}\)$,
$\Da T_{kl} = e^{-K/2}\Da Z^iA_{kli}$.  For general dilaton
couplings, $L_3$ contains the additional term $
{1\over2}f^ie^{-K}\A^jR^{\;\;k\;\;l}_{i\;\;j} A_{kl}\cbW$ which
vanishes in the model considered here since $A_{ss} = 0$.} 
(\ref{dterm}): 
\beq \phi_3 = {1\over2}R^{\alpha k\;\;l}_{\;\;\;\;\alpha}
R^{\dotb}_{\;\; k\dotb l} + 
\(R^{\alpha k\;\;l}_{\;\;\;\;\alpha}e^{-K/2}A_{kl} + {\rm
h.c.}\).\label{l3} \eeq
Cancellation of this term and of the
logarithmic divergence in the renormalization of the K\"ahler
potential in (\ref{li}) require PV chiral superfields $Z^I$ with
nonvanishing $K_{IJ}$, and with superpotential couplings to the light
chiral multiplets.  The part of $K_R$ that depends on the gauge
couplings of the light fields is canceled by superpotential couplings
of the PV fields $\Phi^a$ to the $Z^i$ and to PV chiral fields $Y_I$
that transform according to the gauge group representation that is
conjugate to the light matter representation.  These couplings are
given explicitly in Section 3, slightly modified with respect to those
adopted in I, as required by the presence of the dilaton.  The
superfield form of the operator $L'_0$ is
\bea L'_0 &=& L(\phi'_0) + L(\Phi'_0), \quad \phi'_0 
= \phi_{WK} - 4\phi_{Wk}
- 2\phi_{\cW^a_b} - 4\phi_{\cW}, \nonumber \\ \Phi'_0 &=& 
12\Wc_aT^a_\alpha, \quad T^a_\alpha = - 
{1\over8}\(\bD^2 - 8\R\)\(x^{-1}\D^a\hat{f}_{\alpha}\),
\quad f_i = {\pp f\over \pp Z^i}. \label{lp0}\eea
This term and the remaining contributions to
$L_{0,\chi,g}$ are canceled by the introduction of massive Abelian
gauge fields, some of which couple to the light Yang-Mills fields
through a nontrivial gauge kinetic function, as described in I.
The superfield structure of $L'_g$ is less transparent.  It
is equivalent up to terms that vanish on shell to linear combinations
of the the generic operators introduced above and D-terms that involve
supergravity superfields: $\phi = 
G_{\alpha\dot{\beta}}K_{s\s}\D^\alpha S\D^{\dot{\beta}}\S,\ldots$.
As shown in Appendix C, this term must be exactly canceled by 
PV Abelian gauge multiplets that couple to the dilaton.  

\subsection{PV regularization with a dilaton}

The ultraviolet divergent one-loop corrections to supergravity were
calculated~\cite{us}--\cite{us2} in the presence of a nontrivial gauge 
kinetic function of the form:  
\beq f_{ab}(Z) = \delta_{ab}f(Z)k_a, \quad f(z) = x+iy\ne {\rm constant}.
\eeq 
In~\cite{pv1} it was shown that the dilaton-induced quadratically 
divergent contribution, given by [$T_\alpha$ is defined as in (\ref{tft})]
\beq \STr H \ni - {2N_Gf_i\f_{\m}\over(f+\f)^2}\(\A^iA^{\m} + 
\D_\mu z^i\D^\mu\z^{\m}\) = - N_G\l\Dc T_\alpha\r, \quad
T_i = D_i\ln(f+ \f),  \label{quad}\eeq
can be regulated by the introduction of $N_G$ additional Pauli-Villars 
chiral multiplets $\pi^\alpha$  with
\beq K(\pi,\bar{\pi}) = \sum_\alpha(f + \f)|\pi^\alpha|^2, \;\;\;\;
W(\pi) = \sum_\alpha \mu_\alpha^\pi(\pi^\alpha)^2, \;\;\;\;
\eta^\pi_\alpha = + 1. \label{pi}\eeq
The expression for the logarthmically divergent loop
corrections~\cite{us2} with an arbitrary holomorphic function $f(Z)$
is very complicated.  Here we consider the much simpler case of the
string dilaton, with the dilaton couplings defined by
(\ref{sdil}). For this model (\ref{quad}) takes the form
\beq \STr H \ni - 2N_G\(M^2 + {\pp_\mu s\pp^\mu \s\over4x^2}\)
= N_G\l\Dc\ka\r \label{hdil},\eeq
and the gravitino mass is equal to the gaugino mass:	
\beq M^2_\lambda = M^2_\psi = M^2 = e^{-K}A_s\A^s.
\label{as2}\eeq
In addition we have 
\beq f + \f = e^{-K(S,\S)} = e^{-k}, \eeq
so instead of introducing the additional PV fields in (\ref{pi}), we need only
modify the K\"ahler potential for the gauge fields $\phi^C$ used 
in~\cite{pv1,pvcan} to regulate gravity loops:
\beq K(\phi^C,\bar{\phi}^C) = \sum_C e^{\alpha_CK + \beta_Ck}|\phi^C|^2, 
\label{mod}\eeq
where the case of canonical gauge kinetic energy, $f(Z) = 1$, is recovered 
for $\beta_C= 0$.  

A term proportional to (\ref{hdil}) is also generated if Abelian gauge PV
superfields couple to the dilaton.  We find that it is this latter
mechanism that must be used in order to cancel the dilaton-dependent
logarithmic divergences that arise from gauge loops.  We will also
need to introduce chiral PV multiplets with a K\"ahler potential of
the form (\ref{mod}), with the constraints (see Appendix C) 
\beq
\sum_C\eta^C\beta_C = \sum_C\eta^C\beta_C\alpha_C = 0
.\label{beta}\eeq

\section{Anomaly-free supergravity}
\hspace{0.8cm}\setcounter{equation}{0}
Here we assume that there are no gauge or mixed gauge-gravitational 
anomalies: Tr$T^a =$ Tr$(\{T_a,T_b\}T_c) = 0$,  where $T_a$ is a generator of 
the gauge group.  This section closely follows I, and the reader is referred to
that paper for the contributions that are unchanged when the dilaton is
included. 

We introduce Pauli-Villars chiral supermultiplets 
$Z_\alpha^I= \tZ^I_\alpha,\hZ_\alpha$,
that transform under the gauge group like 
$Z^i_\alpha$, and $Y_I^\alpha = \tY_I^\alpha,\hY_I^\alpha,$
that transform according to the conjugate
representation, as well as gauge singlets $Y^0,Z^0$, and 
chiral multiplets $\Phi^a_\alpha = \varphi^a_\alpha,\tph^a_\alpha,
\hph^a_\alpha$, that transform 
according to the adjoint representation of the gauge group.  
Additional charged fields $X_\beta^A$ and $U^\beta_A$ transform according 
to the representation $R^a_A$ and its conjugate, respectively, under the gauge 
group factor $\G_a$, and $V^A_\beta$ transforms according to a (pseudo)real 
representation that is traceless and anomaly-free.  Their gauge couplings 
satisfy
\beq \sum_{\beta,A} \eta^A_\beta C^a_A = \sum_i C^a_i \equiv C^a_M,\eeq
where 
\beq \Tr_R\(T^aT^b\) = \delta_{ab} C^a_R, \eeq
which may imply a constraint on the matter representations of the gauge group 
in the light spectrum, as discussed in I. In addition, we introduce gauge 
singlets $\varphi^\gamma$, as well as $U(1)$ gauge supermultiplets 
$W_\gamma = W^0_\gamma,
W^s_\gamma$, with signatures $\eta^0_\gamma,\eta^s_\gamma,$ respectively,
that form massive vector supermultiplets with 
chiral multiplets $Z^{0,s}_\gamma = e^{\theta^{0,s}_\gamma}$ of the same signature and 
$U(1)_\beta$ charge $q_\gamma\delta_{\gamma\beta}$.

For the Pauli-Villars fields we take, for illustrative purposes, 
the K\"ahler potential
\bea K_{PV} &=& \sum_\gamma\[e^{\alpha^\phi_\gamma K + \beta^\phi_\gamma
 k}\phi^\gamma\bph_\gamma + {1\over2}
\nu_\gamma(\theta_\gamma + \bth_\gamma)^2 
+ e^{K/2}\sum_A\(|X_\gamma^A|^2 + |U^\gamma_A|^2 + |V^A_\gamma|^2\)\]
\nonumber \\ & & + \sum_{\alpha,a}\(e^G\varphi_\alpha^a\bv_a^\alpha + 
e^k\hph_\alpha^a\hbp_a^\alpha + \tph_\alpha^a\tbp_a^\alpha \) + 
\sum_{\alpha}\(K_\alpha^Z + K_\alpha^Y\) ,
\nonumber \\ K_\alpha^Z &=& \sum_{I,J=i,j}\[K_{i\bj}Z_\alpha^I\Z_\alpha^{\J} 
+ {b^Z\over2}\(K_{IJ}Z_\alpha^IZ^J_\alpha + {\rm h.c.}\)
\] + |Z_\alpha^0|^2, \nonumber \\ 
K_\alpha^Y &=& \sum_{I,J=i,j} K_Y^{I\J}Y^\alpha_I\Y^\alpha_{\J} - 
a_\alpha^Y\sum_{I=i}\(Y^\alpha_I\Y_\alpha^0\kappa_Y^i + {\rm h.c.}\) + 
|Y^\alpha_0|^2\[1 + \(a^Y_\alpha\)^2\kappa_Y^i\kappa^Y_i\], \nonumber \\ 
K^{I\J}_{\hY} &=& K^{i\bj}, \quad 
K^{I\J}_{\tY} = e^{\alpha_IK+\beta_Ik}\delta^{i\bj}, \quad 
\alpha_{I\ne S} = {1\over2}, \quad \beta_S = -2, \quad \alpha_S = 
\beta_{I\ne S} = 0, \nonumber \\ 
K_{IJ} &=& \pp_i\pp_jK - K_iK_j - {1\over2x}\(f_iK_j + f_jK_i\) -
{1\over2x^2}f_if_j, \quad b^{\tZ} = 1, \quad b^{\hZ} = 0, \nonumber \\ 
\kappa^{\tY}_i &=& - {1\over2x}f_i, \quad \kappa_i^{\hY} = K_i 
+ {1\over2x}f_i, \quad \kappa_Y^i = K^{i\m}\kappa^Y_{\m}, 
\quad a^{\tY}_\alpha = 1,
\quad a_\alpha^{\hY} = a_\alpha, \label{kahl}\eea 
and $K^{i\bj}$ is the inverse metric.  We take the superpotential
\bea W_{PV} &=& W_1 + W_2, \nonumber \\
W_1 &=& \sum_{\alpha,\beta}\[\sum_I\mu^Z_{\alpha\beta} Z^I_\alpha Y^\beta_I + 
\mu_{\alpha\beta}^0Z^0_\alpha Y^\beta_0 + 
\sum_a\mu^\Phi_{\alpha\beta}\Phi^a_\alpha\Phi^a_\beta\] \nonumber \\ & & 
+ {1\over2}\sum_{\gamma}\mu^\phi_\gamma\(\phi^\gamma\)^2 
+ \sum_{A\gamma}\(\mu_\gamma^X 
U_A^\gamma X^A_\gamma + {1\over2}\mu_\gamma^V(V_A^\gamma)^2\) \nonumber \\ 
W_2 &=& \sum_\alpha\[a_\alpha W_i\hZ^I_\alpha\hY^\alpha_0 
+ W\hZ^I_\alpha\hY^\alpha_I +
2g_\alpha\varphi^a_{\alpha + 1}\hY_I^\alpha(T_aZ)^i\] \nonumber \\ & &
+ \sum_\alpha\[{1\over 2}\tZ^I_\alpha\tZ^J_\alpha W_{ij}
+ c_\alpha\tZ^S_\alpha\tY^\alpha_S W\],\label{sup}\eea
where the index $a$ refers to the light gauge degrees of freedom.
Finally, we take for the gauge kinetic functions:
\bea f^{ab} &=& \delta^{ab}\(S + \sum_\alpha h_\alpha f_i\tZ^I_\alpha 
\tY_0^\alpha\), \quad f_s^{a\gamma} = 0, 
\nonumber \\ f^0_{\gamma\beta} &=& \delta_{\gamma\beta},\quad 
f^s_{\gamma\beta} = \delta_{\gamma\beta} S, 
\quad f_0^{a\gamma} = \sum_\beta e^{\gamma\beta}\hph^a_\beta, 
.\label{kin} \eea

The matrices $\mu_{\alpha\beta},d_{\alpha\beta},e_{\alpha\beta}$, are 
nonvanishing only when they couple fields of the same signature.
The parameters $\mu,\nu$, play the role of effective cut-offs. 
The parameters
$a,b,c,d,e,h$, are of order unity, and are chosen to satisfy\footnote{The
contribution to $K'$, Eq. (\ref{pvlog}) below, from the last term in 
(\ref{sup}) differs from that of I, where in (2.5) we set 
$c = -2 - N'_G = 10 + N_G$, by the term $-N_GM^2$ needed to cancel the
$N_GM^2$ term in (\ref{li}).}:
\bea a &=& \sum_\alpha\eta_\alpha^{\hY} a_\alpha^2 = -2,
\quad a'=\sum_\alpha\eta^{\hY}_\alpha a_\alpha^4= +2, \nonumber \\
c &=& \sum_\alpha\eta^{\tZ}_\alpha c^2_\alpha = 5, \quad 
g = \sum_\alpha\eta^{\hY}_\alpha g_\alpha^2 a_\alpha^2 = -1, 
\quad \sum_\alpha\eta^{\hY}_\alpha g_\alpha^2 = 1, \nonumber \\
e &=& {1\over2}\sum_{\alpha,\beta}\eta^{\hph}_\alpha e_{\alpha\beta}^2 
= - 4 = 3e', \quad e' = {1\over4}
\sum_{\alpha\beta\gamma\delta}\eta^{\hph}_\gamma
e_\alpha^\beta e_\beta^\gamma e_\gamma^\delta e^\alpha_\delta,
\nonumber \\ h &=& \sum_\alpha\eta^{\tZ}_\alpha h^2_\alpha = 2, \quad
w = \sum_\alpha\eta^{\tZ}_\alpha h_\alpha c_\alpha = 1.\label{cond0} \eea
The signatures of the chiral PV multiplets satisfy
\bea \sum_\alpha\eta^{\varphi}_\alpha &=& \sum_\alpha\eta^{\hph}_\alpha = 
\sum_\alpha\eta^{\tph}_\alpha = 1,
\quad \eta^\varphi_{1 + \alpha} = \eta^{\hZ}_\alpha, 
\quad \eta^\varphi_1 = +1, 
\quad \eta^U_\alpha = \eta^X_\alpha,  
\nonumber \\  \sum_\alpha\eta^{\tZ}_\alpha &=& -1, \quad
\sum_\alpha\eta^{\hZ}_\alpha = 0, \quad \eta^{\tZ}_\alpha = \eta^{\tY}_\alpha,
\quad \eta^{\hZ}_\alpha = \eta^{\hY}_\alpha, \nonumber \\
\sum_\gamma\eta^0_\gamma &=& - 12, \quad \sum_\gamma\eta^s_\gamma = - N_G,
\quad \sum_\gamma\eta^\theta_\gamma = - 12 - N_G = N'_G,  \label{sigs}\eea
and, from the results of I, we require for the exponents in (\ref{kahl})
\beq \alpha = \sum_C\eta_C\alpha_C = -10, \quad \alpha' = 
\sum_C\eta_C\alpha^2_C = -4, \label{cond}\eeq
where in (\ref{cond}) and throughout this section $\phi^C$ is any chiral PV field 
except $Z,\hY$, and $\alpha_\alpha^{\tY^S} = 0,\;\beta_\alpha^{\tY^S}= -2$.  
The K\"ahler potential for $\varphi_\alpha^a$ assures the K\"ahler
anomaly matching condition for the term quadratic in the Yang-Mills field 
strength, as discussed in I and in Section 4 below, as well as 
the correct form of the
gauge-dependent contribution to the renormalization of the K\"ahler potential.

\subsection{Quadratic divergences}

For the class of supergravity theories considered here, the 
on shell\footnote{Specifically, a contribution proportional to
$r - \l\Dc\Xa\r - 6(\hV + M^2)$, where $r$ is the space-time curvature,
can be removed to one-loop order 
by a scalar field dependent Weyl transfromation.}
quadratically divergent contribution is proportional to
\bea \STr H &=& {1\over2}(3+N_G- N)\l\Dc\Xa\r + \(\hV + M^2\)(7 + 3N_G - N)
\nonumber \\ & & + N_G\l\Dc\ka\r + \l\Dc\Gaa\r ,\eea
where $\Xa = K_\alpha$, {\it etc.} are the chiral superfields defined
in (\ref{tft}).
The contribution of the Pauli-Villars fields to $\STr H$ is
\bea \STr H^{PV} &=& \(3\sum_\gamma\eta^\theta_\gamma - 
\sum_P\eta_P\)\(\hV + M^2\) - {1\over2}
\(\sum_P\eta_P - \sum_\gamma\eta^\theta_\gamma\)\l\Dc\Xa\r 
\nonumber \\ & & + \sum_\gamma\eta_\gamma^s\l\Dc\ka\r
+ \sum_P\eta_P\l\Dc\Gamma^P_{P\alpha}\r, \eea
where $P$ refers to all heavy chiral multiplets: $\phi^P = Z^I,Y_I,\phi^C$.
From (\ref{mod}) we have
\bea \Gamma^I_{I\alpha} = \Gaa, \quad \Gamma^C_{D\alpha} = 
\(\alpha_CX_\alpha + \beta_Ck_\alpha\)\delta^C_D,\label{cgam}\eea  
and we obtain for the contribution from heavy PV modes:
\bea \STr H_{PV} &=&  - {1\over2}\(N' - N'_G -2\alpha\)\l\Dc\Xa\r 
+ \(\hV + M^2\)\(3N'_G - N'\) - \l\Dc\Gaa\r \nonumber \\ & &
+ (\beta + f)\l\Dc\ka\r, \nonumber \\
\beta &=& \sum_C\eta_C\beta_C,  \quad N' = \sum_P\eta_P, \quad N'_G = 
\sum_\gamma\eta^\theta_\gamma, \quad f = \sum_\gamma\eta^s_\gamma.\eea
Using (\ref{cond}), the absence of quadratic divergences requires
\bea N' &=& 3\alpha + 1 - N = - 29 - N, \quad \beta + f = - N_G, 
\nonumber \\ N'_G &=& \alpha - 2 - N_G = -12 - N_G .\eea

As explained in~\cite{pv1,pvcan} the $O(\mu^2)$
contribution to $S_0 + S_1 = \int d^4x\(\L_0 + \L_1\)$ takes the form of a
correction to the K\"ahler potential, once additional finiteness 
contraints on the PV masses have been imposed.  Throughout this section we
set (see Appendix C)
\beq \beta = 0,\quad f = - N_G, \quad \beta' = \sum_C\eta_C\beta^2_C = 2, 
\quad \sum_C\eta_C\alpha_C\beta_C = 0.\label{cond2}\eeq

\subsection{Logarithmic divergences}

The Pauli-Villars contribution to (\ref{li}) is, after an appropriate 
space-time metric redefinition,
\bea \L_{PV} &=& \sqrt{g}\lln\[N'_G L_g - N_GL'_g + N' L_\chi + 
\sum_P\eta_P\(L_1^P + L_2^P\) + L_3^Z + L_{\cW} + eL_e\] 
\nonumber \\ & & \quad + 
\Delta_{K'}\L,\quad \quad K' = \lln e^{-K}\sum_{P,Q}\eta_PA_{PQ}\A^{PQ}.
\label{pvlog}\eea 

Using (\ref{cond0})--(\ref{cond}), (\ref{cond2}) and (\ref{as2}),
the PV contributions found in I are modified to read\footnote{See Appendix
B of I and Appendix B below.}
\bea K' &=& \lln\[- e^{-K}A_{ij}\A^{ij} + 2(a + 1)\hV + (2c + 4 - 2a + N'_G)M^2
+ 4\K^a_a +  8g\D\]\nonumber \\ &=& - \lln\[e^{-K}A_{ij}\A^{ij} + 2\hV + 
(N_G - 6)M^2 - 4\K^a_a - 8\D\],\label{pvkahl}\\
L_{\cW} &=& 6e'L(\phi_{\cW^a_b}) + 2hL(\phi_{\cW}) 
- 2wL_{\cW} = 2eL(\phi_{\cW^a_b}) + 4L(\phi_{\cW}) - 2L'_{\cW},\nonumber \\
L_e &=& L(\Phi_e) + L(\phi_e), \quad \Phi_e = -{1\over6}\Phi'_0,
\nonumber \\ \phi_e &=& \phi_{WK} - \phi_{Wk}
- 4\D - 4\phi_{\cW^a_b}, \label{lws} \\
\sum_P\eta_PL_2^P &=& - L_2 - {2\over3}\alpha L_\alpha, \quad 
\sum_P\eta^PL_1^P = - L_1 + 6L^a_{YM}
+ \alpha'L_\alpha + \beta'L_\beta + L_1^Y,  \nonumber \\
L_1^Y + L_3^Z &=& \(L_1^Y + L_3^Z\)_I - {1\over3}L(\Phi'_0)
- 2L_\beta + 2L'_{\cW} - 8\Delta_{M^2}L \nonumber \\ &=&
- L_3 +4\Delta_{\hV}L + 4\Delta_{M^2}L + 8\Delta_{\D}L - 
{1\over3}L(\Phi'_0) - 2L_\beta + 2L'_{\cW} - 8\Delta_{M^2}L, \nonumber \\ 
L_\beta &=& L(\Phi_\beta), \quad \Phi_\beta = - {1\over2}\kc\ka, \quad
L'_{\cW} = x\(\cW + \cbW\)\(\hV + M^2\),\label{l21s}\eea
where the subscript I refers to the result of I.  The contributions 
$L_\alpha$ and $L_\beta$ follow immediately from Eqs. (\ref{cgam}) and 
(\ref{rho}). $L_\beta$ and $L(\Phi'_0)$ are given explicitly in
Appendix B, Eq. (\ref{kop}).

  In I the logarithmic
divergences were found to cancel\footnote{These operators are the bosonic parts
of D-terms of the form (\ref{dterm}) with: 
$$\(\phi_{\cW}\)_I = - 6\phi_{\cW^a_b}, \quad (\phi_e)_I + 4\D =
\phi_{WK} - 4\phi_{\cW^a_b} = 2\hf_0 - 6\phi_\chi -
2\phi_{\cW^a_b}, $$ from which it follows immediately that the
conditions (2.20) and (2.46) of I give $L + L_{PV} = 0$.  The term
$-3C^a\delta_{ab}\(\cW^{ab} + {\rm H.c.}\) + 4\Delta_DL$ is missing
from the right hand side of the third of Eqs. (2.43) of I, and
$8\D$ should be replaced by $(8-4e)\D$ in the second of those equations.}
with $e=-3$.  Hence we write
\bea eL_e + L_{\cW}(e) &=& - 3\(L_e\)_I + \(L_{\cW}\)_I + L'_e - 2L'_{\cW}
 - 4(3+e)\Delta_{\D}L,
 \quad L'_e = L(\phi'_e) + L(\Phi'_e),\nonumber \\ 
\Phi'_e &=& -{e\over6}\Phi'_0, \quad \phi'_e = 4\phi_{\cW} + (3 + e)\(\phi_{WK}
- 2\phi_{\cW^a_b}\) - \phi_{Wk}. \eea
The renormalization of the K\"ahler potential is now finite if $e = -4$.
Complete cancellation of the ultra-violet divergences then requires,
once the conditions (\ref{cond}), (\ref{cond2}) are imposed,
\beq L'_0 + L'_e - {1\over3}L(\Phi'_0) = 0, \eeq
which is achieved for $e= -4$.

\subsection{Soft supersymmetry breaking terms}
Pauli-Villars regularization can be used to calculate one-loop
contributions to soft supersymmetry breaking.  The calculation of
gaugino masses has been given in~\cite{gnw} for string-derived
supergravity with the dilaton in a linear supermultiplet, and
including a Green-Schwarz (GS) term.  These include the ``anomaly
mediated'' contribution~\cite{rs,hit} as well as additional 
model-dependent contributions.  A general analysis of soft
supersymmetry breaking terms in this class of models will be given
elsewhere~\cite{bgn}.  As an example, we calculate
here the one-loop induced A-term for supergravity theories with matter 
in chiral supermultiplets.  To obtain this contribution we take constant
background fields, and the effective one-loop potential is given
simply by 
\bea\L &=& {i\over2}\int{d^4p\over(2\pi)^4}\STr\ln\eta\(p^2 - m^2 - H\)
= -\const\STr\eta\Bigg[\(hm^2 + {1\over2}g^2\)\ln(m^2) \nonumber \\ & & 
\qquad\qquad + 
{1\over2}h^2\ln\({m^2\over\mu^2}\) + {3hg^2\over6m^2} - {g^4\over24m^4} +
O({1\over m^2})\Bigg],\label{lsoft} \eea 
where $\mu$ is the normalization scale, and
$h+g$ is the effective field-dependent squared mass with
the PV mass term removed: 
\beq
H_{PV} = H + m^2, \quad H = h + g, \quad h \sim m^0, \quad g\sim
m^1. \eeq 
We dropped a term $-{1\over6}r\STr H$ in the integrand; in the constant
background field approximation $r\to V$ after a Weyl transformation. 
Assuming $<V>=0$, terms proportional to $V$ can at most contribute
small corrections to soft terms already present at tree level.
The second equality in Eq. (\ref{lsoft}) is schematic: $[H,m^2]\ne 0$
in general.  

Soft terms are generated by the PV fields $\Phi^A = \tZ^I,\hY_I,\varphi^a$
that govern the wave function renormalization
through the dimension three operators in $W_2$, Eq. (\ref{sup}).
We denote by $\Phi^\alpha = \tY^I,\hZ^I,\varphi'^a$, respectively,  
the fields to which to which they couple in $W_1$:
\beq W_1(\Phi^A,\Phi^\alpha) = \sum_{A=\alpha}\mu_A\Phi^A\Phi^\alpha.\eeq
Setting
\beq K_{A\A} = h_A(z), \quad K_{\alpha\bar{\alpha}} = h_\alpha(z),\eeq
we have
\beq m^2_A = m^2_\alpha = f_A\mu_A^2, \quad f_A = e^Kg^{-1}_Ag^{-1}_\alpha. \eeq
The first two terms in Eq. (\ref{lsoft}) are the shift in the potential due to the
shift $\delta K$ in the K\"ahler potential.  The first term,
proportional to $m^2$, corresponds~\cite{pv1} to $\delta K = \sum_Pc_Pm_P^2,\;
c_P=$ constant. They contribute 
A-terms and scalar masses proportional to those already
contained in the tree potential, with coefficients suppressed by the
factor $1/32\pi^2$ ($m^2\sim 1$ in reduced Planck units), and we 
neglect them.

From the general matrix elements evaluated in Appendix C of~\cite{us2}, 
assuming D-terms vanish, dropping derivatives, space-time curvature and
gauge fields, we have 
\bea  (H^\chi)^A_B &=& = h^A_B = e^{-K}A_{AB}\A^{AB}, \quad 
(H^\chi)^\alpha_\beta = 0,
\nonumber \\ (H^\chi)^\alpha_D &=& K^{\alpha\bbet}\mu_BK^{\bB C}A_{CD} = 
g^\alpha_D = e^{-K}f_A\mu_AA_{AD} ,
\nonumber \\  (H^\chi)^A_\beta &=& \A^{AB}\mu_B = g^A_\beta, \quad
(H^\phi)^P_Q = (H^\chi)^P_Q + \delta^P_Q\(\hV + M^2\), \eea
for fermions $\chi$ and scalars $\phi$, respectively. For the reasons
given above we can neglect the $\hV$ 
term, and terms containing only powers of $H^\chi$ cancel 
in the supertrace.  The $M^2$ term does not contribute in leading
order to the A-terms, so they get contributions  only
from the scalar trace terms that have factors of $H^\phi_{PQ}$:
\bea H^\phi_{AB} &=& h_{AB} = e^{-K}\(\A^iD_iA_{AB} - A_{AB}\A\), \quad
h_{\alpha\beta}= 0, \nonumber \\ 
H^\phi_{A\beta} &=& g_{A\beta} = 
e^{-K}\A^iD_i\(e^KW_{A\beta}\) - \A W_{A\beta}
= - \delta_{A\beta}\mu_A\(\A - \A^i\pp_if_A\).\eea
Taking into account the fact that $[H,m^2]\ne 0$ in the integral 
Eq. (\ref{lsoft}), we have on the right hand side:
\bea \Tr h^2\ln(m^2/\mu^2)&\to& 2\sum_{AB}\eta_A\(h^A_Bh^B_A + 
h_{AB}h^{AB}\)\[q(m^2_A,m^2_B) - \ln\mu^2\], \nonumber
\\ q(m^2_A,m^2_B) &=& {m^2_A\ln(m^2_A/\mu^2) -
m^2_B\ln(m^2_B/\mu^2)\over m^2_A - m^2_B} - 1,\label{hln}\eea
and
\bea \Tr{3hg^2\over6m^2} &\to& \sum_{AB}\eta_Ah^A_{\bB}\[g^{\bB}_\beta
g^\beta_A{\pp\over\pp m^2_B}q(m^2_A,m^2_B) + 
g^{\bB}_{\bar{\alpha}}g^{\bar{\alpha}}_A{\pp\over\pp m^2_A}
q(m^2_A,m^2_B)\] + {\rm h.c.} \nonumber \\ &=& - e^{-3K/2}
\sum_{AB}\eta_A\bigg[\(m_{\tG} + \bF^{\m}\pp_{\m}\ln f_B\)
m^2_B{\pp\over\pp m^2_B}q(m^2_A,m^2_B) \nonumber \\ & & \quad
+ \(m_{\tG} + \bF^{\m}\pp_{\m}\ln f_A\)m^2_A{\pp\over\pp m^2_A}
q(m^2_A,m^2_B)\bigg]\eta_A\A^i(D_iA_{AB})\A^{AB}
\nonumber \\ & & \quad + {\rm h.c.} + \cdots ,\label{trh3}\eea
where $\bF^{\m} = - e^{-K/2}A^{\m}$ is the auxiliary field of the
superfield $\Z^{\m}$. In (\ref{trh3}) we have explicitly retained
only contributions to A-terms (and ``B-terms'').  Scalar masses get
contributions from additional terms in (\ref{trh3}) as well as from
Tr$g^4/24m^2$ in (\ref{lsoft}). 
The one-loop corrected scalar kinetic term is
\bea \L_{KE} &=& \D_\mu\z^i\D^\mu\z^{\m}\(K_{i\m} + \delta K_{i\m}\)
= \D_\mu\z^i\D^\mu\z^{\m}(Z^{1\over2})^j_iK_{j\n}(Z^{1\over2})^{\n}_{\m},
\nonumber \\ (Z^{1\over2})^j_i &=& \delta^i_j + 
{1\over2}K^{j\n}\delta K_{i\n}, \nonumber \\ 
\delta K &=& -\const e^{-K}\sum_{AB}\eta_A\A^{AB}A_{AB}
\[q(m^2_A,m^2_B) - \ln\mu^2\],\eea
where $z_R$ is the renormalized field, and the matrix-valued anomalous
dimension is
\bea \gamma^j_i &=& K^{j\n}D_{\n}D_i{\pp\over\pp\mu^2}\delta K = 
\const D^jD_i(e^{-K}\sum_{AB}\eta_A\A^{AB}A_{AB}) =  
e^{-K}\sum_{AB}\eta_A\A^{jAB}A_{iAB} + \cdots,\nonumber \\
&=& \const\[e^KW_{ikl}\W^{jkl} - 
4g^2(T^a\phi)^iK_{j\m}(T_a\bph)^{\m}\] + \ldots.\eea
where the ellipses represent higher order terms. 
To evaluate the A-terms we expand Eqs. (\ref{hln}) and (\ref{trh3}) 
in terms of the light gauge-charged fields $\phi^i$. For  example we have 
\bea \STr h^2 &=& 2\(h_{AB}h^{AB} + h^A_Bh^B_A\)
\nonumber \\ &=& - 2m_{\tG}e^{-3K/2}\A^i(D_iA_{AC})\A^{AC} + {\rm h.c.}
+ \cdots \nonumber \\
&=& - 2m_{\tG}e^{-K/2}\A^i\phi^{\m}K_{j\m}\gamma^j_i + {\rm h.c.}
+ \cdots.\eea
If at tree level we have
\beq K_{i\m} = h_i(z)\delta_{im} + O(|\phi|^2), \quad A_i = 
e^K\(c_{ijk}\phi^j\phi^k + \mu_{ij}\phi^j\) + O|\phi|^3, \eeq
using
\beq \(m^2_B{\pp\over\pp m^2_B} + m^2_A{\pp\over\pp m^2_A}\)q(m^2_A,m^2_B) = 1, \label{h3}\eeq
we get a one-loop contribution to the A-term
\bea \L^1_A &=& \const e^{-K/2}\phi^{\m}\A^iK_{j\m}\gamma^j_i
\[m_{\tG}^2\(1 + \ln(m_{ij}^2/\mu^2)\) + \bF^{\p}\pp_{\p}\ln m^2_{ij}\] + 
{\rm h.c.} \nonumber \\  
&=& \sum_{ijk}e^{-K/2}h^{-1}_j\phi^i_R\phi^k_R
\[\sum_l\phi^l_Rc_{jkl}(h_i/h_kh_l)^{1\over2} + 
m_{jk}(h_i/h_k)^{1\over2}\]\gamma^i_j\times \nonumber \\ & & \qquad \qquad
\[m_{\tG}\(1 + \ln(m_{ij}^2/\mu^2)\) + F^{\m}\pp_{\m}\ln f_{ij} 
\] + {\rm h.c.} + \cdots ,\nonumber \\ \ln m^2_{ij} &=& 
\const\sum_{AB}q(m^2_A,m^2_B)D^jD_i(e^{-K}\A^{AB}A_{AB})/\gamma^j_i
,\nonumber \\ \pp_{\m}\ln f_{ij} &=& \const\sum_{AB}
\[\pp_{\m}q(m^2_A,m^2_B)\]D^jD_i(e^{-K}\A^{AB}A_{AB})/\gamma^j_i.\eea
Note that the term linear in $\phi$
in $A_i$ can arise from a quadratic term in the superpotential or in the
K\"ahler potential; the relation between the corresponding supermultiplet
mass and the one-loop induced ``B-term'' is the same in both cases.
If $m^2_{ij} = \mu^2$ and one assumes canonical kinetic energy for
both the light fields and the PV fields, $\L_A$ reduces to the ``anomaly
mediated'' term found in~\cite{hit}.
The contributions that depend explicitly on the PV masses are contained in
the component field expression of the superfield operator (\ref{delK})
that determines the renormalization of the K\"ahler potential. 
The term proportional to $\ln(m^2_{ij}/\mu^2)$ is not negligible if
the scale of supersymmetry breaking is significantly below the Planck
scale.  A further model dependence is in the $\pp_{\m}ln f_{ij}$ terms.

In contrast to the case of gaugino masses studied in~\cite{gnw}, the
one-loop corrections to the soft terms in the scalar potential are
sensitive to the details of the Pauli-Villars regularization.  In the
gaugino mass case, the PV squared mass matrix commutes with relevant (gauge
superfield dependent) matrix elements.  The regulator masses appear
only through the $\ln m^2$ term, averaged over all charged PV fields,
and only the field dependent part $f_P(z)$ of $m_P^2 = f_P(z)\mu_P^2$
contributes to gaugino masses.  The field dependence ({i.e.}, the
dependence on fields that do not vanish in the vacuum, such as the
dilaton and moduli) on this ``average'' $\ln m^2$ is completely fixed
in terms of the field-dependence of the light field K\"ahler metrics.
Both the requirements of finiteness discussed in Section 3 above and
the supersymmetry of the K\"ahler anomaly~\cite{tom} uniquely
determine the field dependence of $\ln m^2$ once the tree-level theory
(including possible couplings of charged matter to a GS term) is
specified.  However, only a subset of charged PV
fields contribute to the renormalization of the K\"ahler potential.
While the K\"ahler metrics of the fields $\Phi^A$ that appear in $W_2$
is determined by the finiteness requirement, the metrics of the fields
$\Phi^\alpha$ to which they couple in $W_1$ is arbitrary.  Since the
associated K\"ahler anomaly is a D-term, it is supersymmetric by
itself and there is no constraint analogous to the conformal/chiral
anomaly matching in the case of gauge field renormalization with
an F-term anomaly.  As a consequence the ``non-universal''
terms appearing in $\L^1_{soft}$ cannot be determined precisely in the
absence of a detailed theory of Planck scale physics.  In the
following sections we give examples in which the PV masses that
contribute to $\L^1_{soft}$ are field independent.

\section{String-derived supergravity and T-duality}
\hspace{0.8cm}\setcounter{equation}{0} Effective field theories from
superstring compactifications are perturbatively invariant~\cite{mod}
under an $SL(2,Z)$ group (T-duality) of transformations on the chiral
superfields $Z\to Z'(Z)$, which is a subgroup of a continuous
$SL(2,R)$ group, itself a symmetry of the classical Lagrangian. Here
we will refer to both groups as modular transformations. They effect a
K\"ahler transformation: \bea K(Z,\Z) &\to& K(Z',\Z') = K'(Z,\Z) =
K(Z,\Z) + F(Z) + \bF(\Z), \nonumber \\ W(Z) &\to& W(Z') = W'(Z) =
e^{-F(Z)}W(Z), \label{kahlt}\eea and therefore leave the classical
Lagrangian invariant.  Because (\ref{kahlt}) includes phase
transformations on chiral fermions, the symmetry is anomalous at the
quantum level.  Ungauged nonlinear $\sigma$-models were considered in
I, where it was shown that, while the PV K\"ahler potential can be
chosen to be invariant under (\ref{kahlt}), regularization of the
theory with invariant PV masses requires constraints on the light
spectrum.  Moreover, for gauged $\sigma$-models, invariant
regularization does not appear to be possible for any choice of
spectrum.  Indeed, in supergravity theories obtained from orbifold
compactifications of string theory, the (weighted average) masses of
gauge nonsinglet PV chiral multiplets are fixed~\cite{tom} by matching
field theory and string theory loop corrections to the
moduli-Yang-Mills couplings, and cannot all be invariant under
T-duality transformations.

Specifically, we consider a class of orbifold compactifications with,
in addition to the dilaton, the chiral superfields $Z^p = T^i,\Phi^p$, where 
$T^i,\; i=1,2,3,$  are the untwisted moduli, and the K\"ahler potential
\bea G &=& \sum_ig^i + e^{g^p}|\Phi^p|^2 + O\(|\Phi^p|^4\), \quad 
\nonumber \\ g^p &=& \sum_iq^p_ig^i, \quad 
g^i = - \ln\(T^i + \T^{\ibar}\). \eea
The modular transformation
\bea T^i &\to& T'^i = {aT^i - ib\over icT^i + d}, \quad S \to S' = S, 
\quad ad - bc = 1,\nonumber \\
\Phi^p &\to& \Phi'^p = e^{-q^i_p F^i}\Phi^p, \quad F^i = \ln\(icT^i + d\),
\label{modt} \eea
where $q_p^i$ are the modular weights of $\Phi^p$, effects the K\"ahler 
transformation (\ref{kahlt}) with
\beq F(Z) = \sum_i F^i(T^i). \label{ft}\eeq

Setting to zero the gauge-charged background
fields, the one-loop corrected Lagrangian contains the term\footnote{The sign
of this term in (3.3)--(3.7) of I is incorrect}:
\bea \L_1 &\ni& {1\over64\pi^2}\sum_a F_a^{\mu\nu}F^a_{\mu\nu}\sum_\alpha\eta_\alpha\Tr
\(C^\phi_a\ln M^2\)_\alpha,
\nonumber \\ (M^2)^P_Q &=& e^KK^{P\bM}\mu_{\bM\bN}K^{\bN R}\mu_{RQ},\quad
(C^\phi_a)^P_Q = \delta^P_QC^P_a, \label{mass}\eea
and $C^P_a = (\Tr T^2_a)_P$ is the eigenvalue of the quadratic Casimir 
operator on $\phi^P$.  Since the parameters $\mu_{PQ}$ of the superpotential
(\ref{sup}) and -- for vanishing gauge-charged background fields -- the
elements $K_{P\bM}$ of the metric connect only fields $\phi^P$ with the same 
values of $C^P_a$, we have
\beq \sum_\alpha\eta_\alpha\Tr\(C^\phi_a\ln M^2\)_\alpha = 
\sum_P\eta^PC^P_a\Tr\ln M^2_P = 
\sum_P\eta^PC^P_a\ln\Det M^2_P.\eeq
With the choice of K\"ahler potential (\ref{kahl}) we have, for $P,M\ne T^I,S$:
\bea K_Z^{P\bM} &=& \delta^{PM}e^{q^i_Pg^i}, \quad K_{\hY}^{P\bM} = 
\delta^{PM}e^{-q^i_Pg^i}, \quad K_{X,U,V,\tY}^{P\bM} = e^{K/2}\delta^{PM},
\nonumber \\ M^2_{X,U,V} &=& \mu^2_{X,U,V},\quad 
M^2_{\hZ,\hY} = e^K\mu^2_{\hZ,\hY},\quad \Det M^2_{\Phi^a} = 
e^K\Det\mu^2_{\Phi^a}, \nonumber \\
M^2_{\tZ,\tY} &=& e^{{1\over2}(K - 2\sum_iq^ig^i)}\mu^2_{\tZ,\tY}, \quad 
q^i = {\rm diag}(q^i_{p_1},\cdots,q^i_{P_{N-4}}). \eea
Then using the constraints (\ref{sigs}) we obtain
\beq
\L_1 \ni {1\over64\pi^2}\sum_a F_a^{\mu\nu}F^a_{\mu\nu}\[\sum_P\eta^PC^P_a\ln\Det\mu^2_P
- \sum_pC^a_p\(K - 2\sum_iq^i_pg^i\) + C^aK\].\label{gtanom}\eeq
As is well known~\cite{dixon}--\cite{linear},~\cite{tom}, invariance
under (\ref{modt}) is restored by the GS
mechanism; the K\"ahler potential of the dilaton\footnote{The K\"ahler
potential $k$ no longer satisfies $\rho_{ij} = a_i = 0$, in the
notation of~\cite{us2}, resulting in additional contributions to the
loop corrections.  However the modification of $k$ is of one-loop
order, and hence the corresponding one-loop corrections are of
two-loop order, which we do not consider here.} and its modular
transformation property are modified to read 
\beq k = - \ln\(S + \S + {C_{E_8}\over8\pi^2}G\), \quad S' = S - 
{C_{E_8}\over8\pi^2}F,\label{gsterm}\eeq 
so that the variation of $\L_1$, and of model-dependent threshold
corrections, are canceled by a variation in the tree-level coupling
of the dilaton to the Yang-Mills fields.  The contribution in (\ref{gtanom})
satisfies the string matching condition~\cite{tom} when
the Green-Schwarz term and the string-loop threshold corrections are 
included.  Threshold corrections~\cite{dixon,uth} can be 
included as moduli-dependent terms in the PV superpotential $W_1: 
\mu_P = \mu_P(T^i)$.

In order to achieve full perturbative modular invariance, we must
investigate more completely the anomaly structure of the one-loop
corrected effective theory, including gauge nonsinglet background
fields. Supersymmetry relates conformal anomalies, associated with
logarithmic divergences, to chiral anomalies that arise from linearly
divergent integrals in quantum corrections to the low energy effective
theory.  When the theory is regulated in such a way that all integrals
are finite, there are strictly speaking no anomalies, but a
corresponding noninvariance of the quantum corrected theory results
from the noninvariance of the regulator masses. For example, only
light quark loops contribute to the chiral anomaly that permits
neutral pion decay; the anomaly from heavy quark loops is exactly
canceled by the explicit chiral symmetry breaking due to the quark
mass term.  The contribution of a PV quark with negative signature has
the opposite sign; it's anomaly cancels the light quark anomaly and
one is left with the explicit breaking term that exactly reproduces
the light quark anomaly.

Provided we can define modular transformations on the PV fields such
that $K_{PV}$ is invariant and $W_2$ is covariant ($W\to e^{-F}W$),
the noninvariance of the regulated one-loop Lagrangian will arise
solely from the noncovariance of $W_1$ which governs the PV mass-matrix
$M_{PV}$.  The K\"ahler potential for the $\theta_\gamma$ in (\ref{kahl})
is modular invariant provided the chiral superfields $\theta'_\gamma 
= \theta_\gamma$ under (\ref{kahlt}).  
In addition, if we take for the $\Phi^a$ mass term in (\ref{sup}) 
\beq W_1(\Phi^a) = \sum_{\alpha,a}
\[\mu^\varphi_\alpha\varphi^a_\alpha\hph^a_\alpha + {1\over2}
\mu^{\tph}_\alpha\tph^a_\alpha\tph^a_\alpha\] ,\label{phiw1}\eeq
the superpotential for chiral fields $\hph^a$ with dilaton-like couplings 
is modular covariant.  Then the one loop action can be written as 
\beq \L_1 = \L_{inv} + \L_\chi, \quad \L_\chi = {i\over2}\STr\ln\[D^2 +
H(M_{PV})\]_\chi + T_-(M_{PV}),\label{even}\eeq 
where $\L_{inv}$
is modular invariant and $\L_\chi$ contains only chiral supermultiplet
loop contributions.  As a result the masses and covariant
derivatives appearing in the noninvariant contribution contain no
Dirac matrices except in the spin connection, and their contributions
are straightforward to evaluate.

As shown in I, under a transformation on the PV
fields that leaves the tree Lagrangian and the PV K\"ahler potential 
invariant, with $W_2$ covariant:
\bea \Phi' &=& g\Phi, \quad M'_{PV}(\Phi) = M_{PV}(\Phi') \nonumber\\ 
\L' &=& \L_{inv} + \L_\chi(\tM_{PV}), \quad \tM_{PV} = 
g^{-1}M'_{PV}g,\label{lprime}\eea
because all the operators in the determinants except $M_{PV}$ are covariant.
Therefore the anomalous shift in the Lagrangian is given simply by
\beq \Delta \L_1 = \L_\chi(\tM_{PV}) - \L_\chi(M_{PV}).\label{dell}\eeq
As discussed in I, the quadratically divergent terms may be made invariant
by constraints on the PV mass parameters.  In this paper we consider only
anomalies arising from logarithmic divergences and the associated chiral
anomalies.  As a first step toward the construction of a modular invariant
one-loop effective Lagrangian, we give examples below of regularization 
prescriptions with modular covariant PV couplings except in the PV mass terms.
In addition we choose the mass terms such that the renormalization of the
K\"ahler potential is modular invariant.

\subsection{No-scale supergravity}

First we consider a toy ``superstring-inspired'' model~\cite{wit} with a 
single modulus $T$; the K\"ahler potential and superpotential given by
\beq K = k + G, \quad G = -3\ln\(T + \T - \sum_{p=1}^{N-2}
|\Phi^p|^2\), \quad  W = d_{pqr}\Phi^p\Phi^q\Phi^r.\label{toy}\eeq
The modular transformations are defined by
\bea T &\to& T' = {aT - ib\over icT + d}, \quad S \to S' = S, 
\quad ad - bc = 1,\nonumber \\
\Phi^p &\to& \Phi'^p = e^{-F/3}\Phi^p, \quad F = 3\ln\(icT + d\),
\label{toyt} \eea

To construct a modular invariant PV K\"ahler potential and a modular
covariant superpotential $W_2$, we 
note that if the PV K\"ahler potential and superpotential of (\ref{kahl}) and
(\ref{sup}) are modified by the additional terms
\bea K_{PV} &=& K_{PV}^{(3.3)} + \sum_\alpha\[\rho_\alpha
\sum_{I=i}K_iZ_\alpha^IZ_\alpha^{0} + {1\over2}\rho'_\alpha
\(Z_\alpha^0\)^2 + {\rm h.c.}\]\nonumber \\ 
W_2 &=& W_2^{(3.4)} + \sum_\alpha \[\rho_\alpha \sum_{I=i}
W_iZ_\alpha^IZ_\alpha^{0} - {1\over2}\rho'_\alpha\(Z_\alpha^0\)^2W\].
\label{newpot}\eea
the one-loop corrections are unchanged:
\beq  A^{\tZ}_{I0} = R^{\tZ}_{\n I0\m} = 
A^{\tZ}_{00} = R^{\tZ}_{\n 00\m} = 0. \label{zeros}\eeq
For the K\"ahler potential (\ref{toy}) we have, for $Z^i = T,\Phi^q$
\beq K_i = G_i,\quad \pp_i\pp_j G = {1\over3}G_iG_j, \quad K_{IJ} = 
-{2\over3}G_iG_j, \label{prop}\eeq
and, under (\ref{toyt}),
\bea K'_i &=& {\pp K(Z')\over\pp Z'^i} = N^j_i\(K_j + F_j\), 
\quad K'_{i\m} =N^k_iN^{\n}_{\m}K_{k\n},\nonumber \\
K'_{IJ} &=& N^n_iN^m_j\[K_{NM} - {2\over3}\(F_nK_m + F_mK_n + F_mF_n\)\]
.\label{delk}\eea
In addition,
\bea W'_i &=& e^{-F}N^j_i\(W_j - F_jW\), 
\nonumber \\ W'_{ij} &=& 
{\pp^2 W(Z')\over\pp Z'^i\pp Z'^j} = N^k_i\pp_k\[N^m_je^{-F}
\(W_m - F_mW\)\] \nonumber \\ &=& e^{-F}
N^k_iN^m_j\big[W_{km} -F_kW_m - F_mW_k - \(F_{km} - F_kF_m\)W
\nonumber \\ & & - N^l_n\(W_l - F_l\)\pp_kM^n_m\big.\label{delw}\eea 
Writing the transformation (\ref{toyt}) in the form
\bea Z &=& \pmatrix{\Phi^p\cr T\cr} \to Z'(Z), 
\quad M^i_j = {\pp Z'^i\over\pp Z^j}, \quad
N^j_i = {\pp Z'^j\over\pp Z^i}, \nonumber \\
M &=& \pmatrix{e^{-F/3}\delta^p_q & - {1\over3}F_te^{-F/3}\Phi^q\cr
0 & e^{-2F/3}\cr}, \quad 
N = \pmatrix{e^{F/3}\delta^p_q & {1\over3}F_te^{2F/3}\Phi^q\cr
0 & e^{2F/3}\cr},\label{fin}\eea
and using 
\beq W_p\Phi^p = 3W, \quad F_{ij} = - {1\over3}F_iF_j,\eeq
we obtain
\beq W'_{ij} = N^n_iN^m_j\[W_{mn} - {2\over3}\(F_nW_m + 
F_mW_n - F_mF_nW\)\].\eeq
If we also modify the K\"ahler metric for $\tZ$ to read
\beq  K^{\tZ}_{I\J} = K_{i\bj} + a^2G_iG_{\bj}, \quad 
K^{\tZ}_{I\bar{0}} = aG_i, \quad K^{\tZ}_{0\J} = aG_{\bj}, \eeq 
the metric for $\tZ$ is just the inverse of that for $\hY$ 
(see Appendix A of I), {\it i.e.} its inverse is given by 
\beq K_{\Z}^{I\J} = K^{i\bj}, \quad K_{\Z}^{0\J} = - aG^{\bj}, \quad	
K_{\tZ}^{0\bar{0}} = 1 + a^2G^iG_i. \eeq
Because of (\ref{zeros}), there is no additional 
contribution to $L_3^Z$ or $K'$, but there is now a 
contribution $L_1^{\tZ}$ similar to $L_1^{\hY}$.
We can incorporate this contribution if we change the values of $a_{\tY}$
and $a'_{\tY}$ and the parameters in $W_2(\hY)$. 
Moreover, nothing is changed if we
substitute $W_2(\hY,\hZ)\to W_2(\hY,\tZ)$.  Finally, because of the property
(\ref{prop}), the derivatives of the K\"ahler metric $G_{i\m}$ satisfy:
\bea \Gamma_j^{i\alpha}\Gamma_{i\alpha}^j &=& 
{1\over9}\[\(N+1\)G^\alpha G_\alpha  + G^{i\alpha}_j G^j_{i\alpha}\],
\quad \Gamma^i_{i\alpha} = {N\over3}G_\alpha, 
\nonumber \\ G_\alpha &=& G^i_{i\alpha}, 
\quad G^j_{i\alpha} = -{1\over8}\(\DbDb - 8R\)\(G_i\D_\alpha Z^j\),  \nonumber \\
\Gamma_{i\alpha}^j(T^a)^i_j &=& {1\over3}G^a_\alpha, \quad
G^a_\alpha = -{1\over8}\(\DbDb - 8R\)\[G_i\D_\alpha(T^aZ)^i\],  
\label{toyk}\eea  
while the derivatives of the metrics for $\phi^C$ with 
$\beta_C = -1$, and for $\hY_{P\ne S}$, satisfy [see (2.41) of I]
\bea &&\Gamma_D^{C\alpha}\Gamma_{C\alpha}^D = 
\alpha_C^2G^\alpha G_\alpha , \quad
\Gamma_{D\alpha}^C = \alpha_CG_\alpha\delta^C_D, \nonumber \\ &&
\Gamma_{P\alpha}^Q(T^a)^P_Q = \Gamma_{i\alpha}^j(T^a)^i_j + a^2G^a_\alpha,
 \quad \Gamma_{P\alpha}^P = -\Gamma^i_{i\alpha},\nonumber \\
&&\(\Gamma\)_Q^{P\alpha}\(\Gamma\)_{P\alpha}^Q = 
\Gamma_j^{i\alpha}\Gamma_{i\alpha}^j +
\({2\over3}a^2 + a^4\)
\(G^\alpha G_\alpha  + G^{i\alpha}_j G^j_{i\alpha}\) 
- 2\({1\over3}a^2 + a^4\)G_{i\alpha}Z^i_\alpha, \nonumber \\ &&
G_{i\alpha} = -{1\over8}\(\DbDb - 8R\)G_i\Da G, \quad
Z^i_\alpha = -{1\over8}\(\DbDb - 8R\)\Da Z^i.\label{metr} \eea  
Therefore, since in addition $k_{s\s} = e^{2k}$, 
the contribution of fields with metric 
$K_{i\m}$ can be canceled by an appropriate combination of $\hY,\tZ,\phi$,
provided some $\phi$ are gauge-charged.

As a consequence of the above, the ultraviolet divergences are still canceled 
if we modify (\ref{kahl}), (\ref{sup}), (\ref{cond0}) and
(\ref{sigs}) to read [note that $k^ik_i = 1$, and $G^iG_i$ = 3 is invariant
under the modular transformations (\ref{toyt})]
\bea  K_{PV} &=& \sum_\gamma\[e^{\alpha^\phi_\gamma K + \beta^\phi_\gamma  k}
\phi^\gamma\bph_\gamma + {1\over2}\nu_\gamma(\theta_\gamma + \bth_\gamma)^2\]
+ e^{K/2}\sum_A\(|X_\gamma^A|^2 + |U^\gamma_A|^2 + |V^A_\gamma|^2\)
\nonumber \\ & & + \sum_{\alpha}\[\sum_a\(e^G\varphi_\alpha^a\bv_a^\alpha + 
e^k\hph_\alpha^a\hbp_a^\alpha + \tph_\alpha^a\tbp_a^\alpha\) 
+ e^K\sum_{r=1}^3e^{\beta^r_\alpha k}|\phi^r_\alpha|^2\]\nonumber \\ & & 
+ \sum_{\alpha}\[e^{-2k}|\phi_S^\alpha|^2 + 2|\phi_0^\alpha|^2 - 
e^{-k}\(\bph_{\S}^\alpha\phi_0^\alpha + {\rm h.c.}\)  
+ e^{2k}|\phi^S_\alpha|^2\] \nonumber \\ & & 
+ \sum_{\alpha}\[\sum_{I\ne S}\(e^{G/3}|\phi^I_\alpha|^2 + e^{K/2}
|\phi_I^\alpha|^2\)  + e^{G/3}|\phi^0_\alpha|^2 + K_\alpha^Z + K_\alpha^Y\]
\nonumber \\ K_\alpha^Z &=& e^{\alpha^ZG}\Bigg\{
\sum_{I,J=i,j}\[Z_\alpha^I\Z_\alpha^{\J}\(G_{i\bj}
+ a_\alpha^2G_iG_{\bj}\) - {b\over3}\(G_iG_jZ_\alpha^IZ^J_\alpha 
+ {\rm h.c.}\)\] + |Z_\alpha^0|^2 
\nonumber \\ & & - \[\sum_{I=i}G_iZ^I_\alpha
\({2b\over3a_\alpha}Z^0_\alpha - a_\alpha\Z^0_\alpha\) 
+ {b\over3a^2_\alpha}\(Z^0_\alpha\)^2 
+ {\rm h.c.}\]\Bigg\},\nonumber \\ 
K_\alpha^Y &=& e^{\alpha^YG}\[\sum_{I,J=i,j}G^{i\bj}Y^\alpha_I\Y^\alpha_{\J} - 
a_\alpha\sum_{I=i}\(Y^\alpha_I\Y_\alpha^0G^i + {\rm h.c.}\) + 
|Y^\alpha_0|^2\(1 + 3a_\alpha^2\)\], \nonumber \\ 
{\tilde b} &=& 1, \quad \hat{b} = 0, \quad \alpha^{\tZ} = 
\alpha^{\hY} = 0, \quad \alpha^{\hZ} = \alpha^{\tY} = 1,\label{newk} \\ 
W_1 &=& \sum_\alpha\[\sum_{P\ne S}\({\tilde\mu}_\alpha\tZ^P_\alpha\tY^\alpha_P
+ \hat{\mu}_\alpha\hZ^P_\alpha\hY^\alpha_P\) + 
\sum_{I\ne S}\mu^\phi_\alpha\(\phi^I_\alpha\phi_I^\alpha 
+{1\over2}(\phi^0_\alpha)^2\)\] \nonumber \\ & &
+ \sum_{\alpha}\[\sum_a\(\mu^\varphi_\alpha\varphi^a_\alpha\hph^a_\alpha + 
{1\over2}\mu^{\tph}_\alpha\tph^a_\alpha\tph^a_\alpha\)
+ \sum_r\mu_\alpha^S\phi_\alpha^r\hf^\alpha_r\] 
\nonumber \\ & & + \sum_{A\gamma}\(\mu_\gamma^X 
U_A^\gamma X^A_\gamma + {1\over2}\mu_\gamma^V(V_A^\gamma)^2\) 
+ {1\over2}\sum_{C,D}\mu^\phi_{CD}\phi^C\phi^D,\label{neww1} \\
W_2 &=& \sum_\alpha\lbr{1\over2}W_{ij}\tZ_\alpha^{I}\tZ_\alpha^{J} + 
{1\over3a_\alpha^2}\(\tZ_\alpha^0\)^2W
- {2\over3a_\alpha}W_i\tZ_\alpha^I\tZ_\alpha^0  
+ 2\sum_a\varphi_\alpha^a\hY^\alpha_I(T_aZ)^i\rbr \nonumber \\ & & 
+ \sqrt{2}\sum_{\alpha>1}\tZ^I_\alpha\(\hY_I^\alpha W + \hat{a}_\alpha
W_i\hY^\alpha_0\) + \sum_\alpha
c_\alpha\phi_\alpha^S\phi^\alpha_S W, \label{neww2} \\ 
f^{ab} &=& \delta^{ab}\(s + \sum_\alpha h_\alpha\phi^S_\alpha\phi_0^\alpha
\), \quad f_s^{a\alpha} = 0, 
\nonumber \\ f^0_{\alpha\beta} &=& \delta_{\alpha\beta},\quad 
f^s_{\alpha\beta} = \delta_{\alpha\beta} S, 
\quad f_0^{a\alpha} = \sum_\beta e^{\alpha\beta}\hph^a_\beta, \label{newf} \\ 
{\tilde a} &=& - {1\over6}, \quad {\tilde a}' = {1\over18}, 
\quad \hat{a} = - \hat{a}' = -1, \quad \hat{a}_1 = 0,
\nonumber \\ 
h &=& 2, \quad e = - 4 = 3e', \quad c = 5,  \quad w = 1, \label{newcond} \\
\sum_\alpha\eta^{\varphi}_\alpha &=& \sum_\alpha\eta^{\hph}_\alpha = 
\sum_\alpha\eta^{\tph}_\alpha = \sum_\alpha\eta^{\hZ}_\alpha = -
\sum_\alpha\eta^{\tZ}_\alpha = - \sum_\alpha\eta^r_\alpha = + 1, 
\nonumber \\ \eta^\varphi_\alpha &=& \eta^{\hZ}_\alpha = 
\eta^{\hY}_\alpha, \quad \eta^{\hph}_\alpha = \eta^\varphi_\alpha, \quad
\eta^r_\alpha = \eta^{\phi^{I,0}}_\alpha = \eta^{\phi_{I,0}}_\alpha, 
\quad \eta^U_\alpha = \eta^X_\alpha, \nonumber \\
\eta^{\tY}_\alpha &=& \eta^{\tZ}_\alpha, \quad 
\eta^{\tZ}_{\alpha + 1} = \eta^{\hY}_{\alpha + 1}, \quad 
\eta^{\tZ}_1 = - \eta^{\hY}_1 = -1, \nonumber \\
\sum_\gamma\eta^0_\gamma &=& - 12, \quad \sum_\gamma\eta^s_\gamma = - N_G,
\quad \sum_\gamma\eta^\theta_\gamma = - 12 - N_G = N'_G, \label{neweta}\eea
where $\phi^I,\phi_I$ transform like $Z^I,Y_I,$ respectively, under the gauge
group, $\hf^r = \phi^S,\phi_S,\phi_0,$
and in $W_1$ the sum over $\phi^C$ includes $\phi_{P=I,0}$ but not
$\phi^{P=I,0}$, and $a,a'$ are defined as in (\ref{cond0}).

The metric derivatives for $\tZ$ are the same as for $\hY$ in
(\ref{metr}) except that $(\Gamma^Q_{P\alpha})_Z = - 
(\Gamma^P_{Q\alpha})_Y$, and the derivatives for $X' = \tY,\hZ$ are 
related to those for $X = \tZ,\hY$ by
\bea && \(\Gamma_{X'}\)_{P\alpha}^P = - 
\(\Gamma_{X}\)_{P\alpha}^P + NG_\alpha, 
\quad \(\Gamma_{X'}\)_{P\alpha}^Q(T_a)^P_Q = 
\(\Gamma_{X}\)_{P\alpha}^Q(T_a)^P_Q \nonumber \\
&&\(\Gamma_{X'}\)_Q^{P\alpha}\(\Gamma_{X'}\)_{P\alpha}^Q = 
\(\Gamma_{X}\)_Q^{P\alpha}\(\Gamma_{X}\)_{P\alpha}^Q
+ N\(1 \pm {2\over3}\)G^\alpha G_\alpha .\eea
Since $\hZ$ and $\tY$ have opposite signature, the additional surviving 
contributions are equivalent to that of a set $\phi^C$ with $\alpha = 
\beta = 0, \;\alpha' = \beta' = - \sigma = - 4N/3$.  To cancel this 
contribution we must modify (\ref{cond}) and (\ref{cond2}) to read
\beq f = - N_G, \quad 4 + \alpha' = \beta' = - \sigma = + 4N/3, 
\label{newconst}\eeq
where in the sums defining these quantities
\beq \phi^C = \phi_\gamma,X^A_\gamma,U_A^\gamma,V^A_\gamma,\phi_P^\gamma,
\Phi^a_\gamma. \label{phic}\eeq
In other words [see (\ref{betal}) of Appendix C],
\bea \sum_IL_2^{\phi^I} &=& - L_2, \quad L_2^Y + L_2^Z = 0, \nonumber \\
\sum_IL_1^{\phi^I} + L_1^Y + L_1^Z &=& - L_1 - {4N\over3}\(L_\alpha +
L_\beta + L'_\beta\). \eea
Since $(T_aZ)^iF_i = 0$, 
$K_{PV}$ is invariant and $W_2$ is covariant ($W_2\to e^{-F}W_2$)
under (\ref{toyt}) provided the PV chiral multiplets transform as 
\bea \phi'^C &=& e^{-\alpha_CF}\phi^C, \quad
Y'^\alpha_I = e^{-\alpha^YF}N_i^j\(Y^\alpha_J + a_\alpha F_jY_0\), 
\quad Y'_0 = e^{-\alpha^YF}Y_0\nonumber \\
Z'^J_\alpha &=& e^{-\alpha^ZF}M_i^jZ^I_\alpha, \quad
Z'^0_\alpha = e^{-\alpha^ZF}\(Z^0 - a_\alpha Z^I_\alpha F_i\), 
\label{pvt}\eea
with all other PV superfields invariant.  Note that we have chosen $W_1$ such 
that all masses are covariant for fields that appear in the gauge kinetic
functions $f^{AB}$.  Provided each $\phi^C$ appears in only one term in $W_1$
[{\it i.e.} $\mu_{CC'}\phi^C\phi^{C'}$ or ${1\over2}\mu_C(\phi^C)^2], $
the squared-mass matrix defined in (\ref{mass}) is block diagonal
Thus, for example, if we include a modular
covariant $T$-dependence in the mass terms for some $\phi^P\ne Z,Y$, we have
\bea (M^2_{\tZ})^P_Q &=& (M^2_{\tY})^Q_P = {\tilde\mu}^2\delta^P_Q, \quad
(M^2_{\hZ})^P_Q = (M^2_{\hY})^Q_P = \hat{\mu}^2\delta^P_Q, \nonumber \\
M^2_{\phi^P} &=& M^2_{\phi^{P'}} = 
{\tilde\mu}_{PP'}^2|\eta(it)|^{4b_P}e^{K(1 - \alpha_P - \alpha_{P'})
- k(\beta_P + \beta_{P'})}, \label{m2}\eea
where $\eta(it)$ is the Dedekind function:
$$ \eta(iT'^i) = e^{{1\over2}F^i}\eta(iT^i),$$ and 
\beq \tM^2_{\phi^P} = M'^2_{\phi^P} = e^{(F+\bF)
(1 - \alpha_P - \alpha_{P'} + b_P)}M^2_{\phi^P},
\quad \phi^P\ne Z,Y \label{tm},\label{tm2}\eeq
with $\tM^2 = M^2$ otherwise.  The $\eta(it)$ factor can be
interpreted as a parameterization of string loop threshold corrections,
as mentioned in Section 4.1.  We 
now turn to a more realistic model from string theory.

\subsection{The untwisted sector of orbifold compactifications}
Consider next the classical Lagrangian for the untwisted sector 
of orbifold compactifications with three untwisted moduli.  It is
defined by the K\"ahler potential and the superpotential\footnote{It
is straightforward, but slightly more cumbersome, to generalize the
results to the case of a K\"ahler potential as in (\ref{untwk}) with
$n\to n_i,\; n_i\ne n_j$.}
\bea K &=& k + G, \quad G = G^u, \quad 
W = d_{abc}\Phi^{a1}\Phi^{b2}\Phi^{c3},\nonumber \\
G^u &=& \sum_{i=1}^3\Gi,\quad \Gi = -\ln\(T^i + \T^{\ibar} - \sum_{a=1}^{n}
|\Phi^{ai}|^2\),\label{untwk}\eea
Setting $Z^{p} = \{T^i,\Phi^{(ai)}\}$, we now have the 
properties
\bea \pp_{p}\pp_{q} G &=& \delta_{ij}\Gi_{p}\Gi_{q}, 
\quad K_{pq} = - \sum_{i\ne j}G_{p}G_{q}, \nonumber \\
\sum_pW_{p}G^{p} &=& 0,\quad \sum_a W_{(ai)}\Phi^{(ai)} = W.
\label{mprop}\eea
The Lagrangian is invariant under modular transformations:
\bea G &\to& G' = F + \bF, \quad 
F = \sum_iF^i, \quad F^i = \ln(icT^i + d), \nonumber \\ 
Z &=& \pmatrix{\Phi^p\cr T\cr} \to Z'(Z), 
\quad M^{p}_{q} = {\pp Z'^{p}\over\pp Z^{q}}, \quad
N^{q}_{p} = {\pp Z'^{q}\over\pp Z^{p}}, \nonumber \\
M &=& \delta^i_j\pmatrix{e^{-F^i}\delta^a_b & - F_ie^{-F^i}\Phi^{(ai)}\cr
0 & e^{-2F^i}\cr}, \quad 
N = \delta^i_j\pmatrix{e^{F^i}\delta^a_b & F_ie^{2F^i}\Phi^{(ai)}\cr
0 & e^{2F^i}\cr}, \nonumber \\
F_i &\equiv& F_{t^i} = F^i_{t^i}, 
\quad F_{ij} = - \delta_{ij}F^2_i. \label{fmod}\eea
Properties analogous to (\ref{delk}), (\ref{delw}), (\ref{toyk}) and
(\ref{metr}) are given in Appendix D.
In analogy with the discussion of the  preceding section, we introduce PV superfields 
$Z^{(0I)},Y_{(0I)}$, and modify $K^{Y,Z}$ and $W_2(Z)$ 
in (\ref{newk}) and (\ref{neww2}) to read [here we suppress the index $\alpha$,
and now $\Gii^p\Gi_p = 1$ for fixed $i$ is invariant under (\ref{fmod})]
\bea K^{Z} &=& e^{\alpha^ZG}\sum_{I=1}^3K_I^Z - 
{b_Z\over2}\sum_{I\ne J}\(G_Z^IG_Z^J +
{\rm h.c.}\), \quad K^{Y} = e^{\alpha^YG}\sum_{I=1}^3K_I^Y, \nonumber \\
K_I^Z &=& \sum_{P\bM}Z^{P}\Z^{\bM}\Gi_{p\m} 
+ |aG_Z^I|^2, \quad G^I_Z = \sum_PZ^{P}\Gi_{p} +  a^{-1}Z^{(0I)},
\nonumber \\ K_I^Y &=& \[\sum_{P\bM}Y_{P}\Y_{\bM}\Gii^{p\m} 
 - a\sum_P\(Y_{P}\Y_{(\bar{0}I)}\Gii^{p} + {\rm h.c.}\) 
+ Y_{(0I)}\Y_{(\bar{0}I)}\(1 + a^2\)\], \nonumber \\ 
b_{\tZ} &=& - 2{\tilde a} = 2{\tilde a}' = - \hat{a} = \hat{a}' = 1, 
\quad b_{\hZ} = 0, \label{modk} \\
W_2(\tZ) &=& {1\over2}W_{pq}\tZ^{P}\tZ^{Q} - a^{-1}\sum_{i\ne j}
\tZ^{0J}\(W_{(ai)}\tZ^{(AI)} - {1\over2}a^{-1}W\tZ^{(0I)}\) \nonumber \\ & & 
+ \sqrt{2}\sum_{P,\alpha>1}\tZ^P\(aW_p\hY_0 + \hY_P W\). \eea
In addition we replace the fields $\phi^{I,0},\phi_I,\;I\ne S$ by 
$\phi^{(PI),(P0)},\phi_{(0I)},\;P\ne S$, with K\"ahler potential
\beq K^\phi = \sum_i\[e^{\Gi}\(\sum_P|\phi^{(PI)}|^2 + |\phi^{(0I)}|^2\) + 
e^{K/2}\sum_P|\phi_{(PI)}|^2\].\eeq
The mass terms for theses fields are determined by $W_1$ in (\ref{neww1}) with 
$$Z^P,Y_P \to T^I,Z^{(AI)},T_I,Y_{(AI)}, \quad 
Z^0,Y_0 \to Z^{(0I)},Y_{(0I)}, $$ $$ \phi^{I,0},\phi_I, \;\; 
I\ne S,\;\; \to \phi^{(PI),(0I)},\phi_{(PI)}, \; \;P\ne S, $$
and the sum over $C$ in the definitions of $\alpha,\alpha'$
now includes $\phi_{(PI)}$.
The K\"ahler potential is invariant and $W_2$ is covariant under modular 
transformations provided 
\bea \phi'^C &=& e^{-\alpha_CF}\phi^C, \quad
Y'_{P=T_I,(AI)} = e^{-\alpha^YF}N_{p}^{q}\(Y_{Q} + a F^i_{q}Y_{(0I)}\), 
\nonumber \\ Z'^{Q} &=& e^{-\alpha^ZF}M_{p}^{q}Z^{P}, \quad 
Z'^{(0I)} = e^{-\alpha^ZF}\(Z^{(0I)} - a Z^{P} F^i_{p}\), 
\nonumber \\ Y'_{(0I)} &=& e^{-\alpha^YF}Y_{(0I)},
\quad \phi'^{(NI)} = e^{-F^i}\phi^{(NI)}, \quad N = P,0. \label{pvm}\eea
The renormalization of the K\"ahler potential (\ref{pvkahl}) arises
from $Z,Y,\varphi,\theta,\phi^S,\phi_S,$ contributions and is modular
invariant, since we have chosen the PV couplings such that their
masses are covariant. Writing,  for $\phi^P\ne Z,Y$, 
\bea K(\phi^P,\bph^P) &=& e^{G_P +\beta_Pk}|\phi^P|^2, \quad G_P= \sum_i
\alpha^i_P\Gi, \nonumber \\ W_1(\phi^P,\phi^{P'}) &=& \mu_P
\prod_i[\eta(it^i)]^{2b^i_P}\phi^P\phi^{P'}, \eea
we have
\bea M^2_{\phi^P} &=& M^2_{\phi^{P'}} = 
\mu_P^2\prod_i|\eta(it^i)|^{4b^i_P}e^{K - G_P - G_{P'}
- k(\beta_P + \beta_{P'})}, \label{m2u}\eea
and 
\beq \tM^2_{\phi^P} = M'^2_{\phi^P} = e^{\sum_i(\Fi+\bFi)
(1 - \alpha_P^i - \alpha^i_{P'} + b^i_P)}M^2_{\phi^P},
\quad \phi^P\ne Z,Y .\label{tm2u}\eeq

\subsection{Including the twisted sector}
The K\"ahler potential for orbifolds is not known beyond leading (quadratic)
order in the fields $Z^a\ne S,T$, except for the untwisted sector, 
whose K\"ahler potential (\ref{untwk}) is determined by the metric on the 
compact space.  As a consequence, we cannot determine the 
one-loop effective action for the twisted sector, but we can include twisted
sector loop contributions to the untwisted sector action, provided 
the superpotential contains no terms quadratic in the twisted sector fields. 
The general modular invariant superpotential\footnote{There can be  
additional factors which are holomorphic, modular invariant functions
of the moduli.}
\beq W = \sum_\alpha w_\alpha\prod_{j=1}^3\eta^{-2}(T^j)\prod_a\[
Z^a\prod_{i=1}^3\eta^{2q^a_i}(T^i)\],\eeq
depends on the moduli through the Dedekind $\eta$-function, interpreted as 
arising from string world-sheet instanton effects.  In the absence of these 
effects, which we neglect here, there is no superpotential for twisted sector
fields.  We will set background twisted sector fields to zero, and include
only quantum corrections due to the (modular invariant) quadratic term 
in $Z^a\ne S,T^i,\Phi^{ia}$ in the superpotential:
\bea K &=& k + G^u + \sum_ae^{g^a}|Z^a|^2, \nonumber \\ g^a &=& -\sum_i
q^i_a\ln(T^i+\T^{\ibar}) + f^a[|\Phi^{bi}|^2/(T^i+\T^{\ibar})]
\label{twk}\eea
If $K$ depends on the moduli only through the compact radii, we have
\beq g^a = G^a = \sum_iq^i_aG^i, \quad f_a = -\sum_iq^i_a\ln[1 -\sum_b
|\Phi^{bi}|^2/(T^i+\T^{\ibar})].\label{heis}\eeq
Under a modular transformation 
\bea Z'^a &=& e^{-F^a}Z^a, \quad F^a = \sum_iq^i_aF^i.  \eea 
To regulate the twisted sector contribution, we introduce negative signature
PV fields 
$\Phi^A,\Phi_A$ that transform under the gauge group like $\Phi^a$
and its conjugate, respectively, with K\"ahler potential and superpotential
\beq K_{PV}^T = \sum_A\(e^{g^a}|\Phi^A|^2 + e^{K/2}|\Phi_A|^2\),
\quad W_1^T = \sum_A\prod_i|\eta(it^i)|^{2b^i_A}\mu_A\Phi^A\Phi_A. \eeq
Under (\ref{modt}) we have
\bea \Phi^A &\to& e^{-F^a}\Phi^A, \quad \Phi_A\to e^{-{1\over2}F}\Phi_A,
\nonumber \\ (\tM^A)^2 &=& \tM^2_A 
= e^{\sum_i(\Fi +\bFi)({1\over2} - q^i_a + b^i_A)}M_A^2.\eea
Combining this with (\ref{m2u}), (\ref{tm2u}), 
the one-loop Yang-Mills Lagrangian (\ref{gtanom})
takes the form\footnote{There
are additional dilaton-dependent terms (formally of two-loop order) if
the gauge charged fields couple to the GS term (ref{gsterm}).}
\beq
\L_1 \ni {1\over64\pi^2}\sum_a F_a^{\mu\nu}F^a_{\mu\nu}
\[\sum_P\eta^PC^P_ab^i_P\ln|\eta(it^i)|^4
- \sum_pC^a_p\(K - 2\sum_iq^i_pg^i\) + C^aK\],\label{orbym}\eeq
where the sum over $P$ now included the twisted sector fields.
The first term in (\ref{orbym}) correctly reproduces the threshold
effects (neglecting the universal, modular invariant term~\cite{uth})
provided
\beq  b^i_a = \sum_P\eta^PC^P_ab^i_P = C_{E_8} + \sum_pC^a_p\(1 - 
2q^i_p\) - C^a. \label{thr}\eeq
Then the variation in (\ref{orbym}) is cancelled by the variation in
the classical Yang-Mills Lagrangian due to the transformation property
(\ref{gsterm}) of the dilaton.

\section{Anomalous $U(1)$}
\hspace{0.8cm}\setcounter{equation}{0} 
The modifications needed for
regulating one-loop supergravity in the presence of an anomalous
$U(1)$ gauge group $\G_X$ are described in detail in I.  The light
matter loops generate a quadratically divergent term proportional to
$2x^{-1}D_X\Tr T_X$ and logarithmic divergences proportional to $\Tr
T_X$ associated with the operators $\Phi_{1,2}$ in (\ref{rho}).  To
regulate these terms we must introduce PV chiral multiplets $\phi^P$
with superpotential terms that are not invariant under \ux.
As discussed in I, in order for the superpotential to remain
holomorphic under a \ux\, gauge transformation, we require the
transformation properties 
\bea {\cal A}^X_M&\to&{\cal A}^X_M -
g^{-1}D_Mg, \quad\vx\to\vx' = \vx + {1\over2}\(\Lambda + \bL\),
\nonumber \\ Z^i&\to&g^{q^i_X}Z^i, \quad Z^I\to g^{-q^i_X\Lambda}Z^I,
\quad g = (g^{\dag})^{-1} = e^{{1\over2}(\bar{\Lambda} - \Lambda)}.
\label{gauge}\eea 
The chiral Yang-Mills superfield $W^\alpha$ is
obtained as a component~\cite{bggm} of the two-form ${\cal F}_{MN}$,
which is the super-curl of the Yang-Mills one-form potential ${\cal A}_M$,
and is also the chiral projection of the commonly used Yang-Mills
superfield potential $V_X$: $W_\alpha = - {1\over4}\(\DbDb-8R\)\Dc
V_X.$ While the light fields are defined to be covariantly
chiral~\cite{bggm} under \ux, the \ux-charged PV fields are covariantly
chiral only with respect to the nonanomalous gauge group; 
their invariant superpotential takes the form 
\beq
K_{PV}(|\phi^P|^2) = e^{g^P(Z) + 2q_X^P\vx}|\phi^P|^2.\eeq 

\subsection{General supergravity}
If we assume that the \ux\, generator commutes with the K\"ahler
metric in the general supergravity model of Section 2, we can simply
assign zero \ux,\ charge to $X_\gamma,U_\gamma,V_\gamma$, and to
$\hY_I^\alpha$ for a set of values $\alpha = \a0$ with
$\sum_\alpha\eta_{\a0} = -1$. \ux\, gauge invariance of $K_{PV}$ and
$W_2$ as defined in Eq. (\ref{sup}) requires $a_{\a0} = g_{\a0}$. We
must also remove $\hY_I^{\a0},\hZ^I_{\a0}$ as well as a pair with
$\alpha\ne\a0$ and net positive signature from the second term in
$W_2$, Eq. (\ref{sup}).  With this choice the linear divergences associated with
the \ux\, anomaly are canceled.  The chiral anomaly reappears due to
the noninvariance of the mass terms coupling the $\hY^{\a0}$ to fields
$\hZ^I_{\a0}$ with the same \ux\, charge as $Z^i$, and forms a
supersymmetric F-term with the chiral anomaly.  Note that the
renormalization of the K\"ahler potential is \ux\, invariant in this
general case, since $\hY_{\a0},\hZ^I_{\a0}$ do not appear in $W_2$.

\subsection{Orbifold compactification}
In this case we cannot impose the condition that the K\"ahler metric
commutes with the \ux\, generator, but with an appropriate choice of
PV \ux\, charges and superpotential, the \ux\, generator does commute
with the K\"ahler metric for PV fields with PV masses that are not
\ux\, covariant.  For the untwisted sector of the orbifold model of
Sections 4.2-3, we have
\bea \Gamma^{(pi)}_{(qi)\alpha}(T_X)^{(pi)}_{(qi)} &=& \Gia(T_X)^{(pi)}_{(pi)} + 
\G^X_\alpha = \Gamma^{(PI)}_{(QI)\alpha}(T_X)^{(PI)}_{(QI)} + \G^X_\alpha, 
\nonumber \\ \Gamma^b_{a\alpha}(T_X)^a_b &=& g^a_\alpha(T_X)^a_a =
\Gamma^B_{A\alpha}(T_X)^A_B. \eea
The contribution from $\G^X_\alpha$, which is defined in (\ref{toyk}), is 
canceled as before provided $q^{Z^{(AI)}}_X = - q^{Y_{(AI)}}_X = 
q^{Z^{(ai)}}_X$ and to cancel the new contributions, we assign
$U(1)_X$ charge to $\phi^{(AI)}$: $q^{(AI)}_X = q^{(aI)}_X$ and to 
$\phi^C$: $q_X = \sum_C\eta_C\alpha_Cq^C_X = - 2,\;\sum_C\eta_C
\beta_Cq^C_X = \sum_C\eta_Cq^C_X = 0$,
where $q_X$ is chosen to cancel the contribution from the last term in
(D.4) of Appendix 4.  We also require $q^A_X = q^a_X$ for the PV 
regulator fields for the twisted sector.  With these choices, the
renormalization of the K\"ahler potential is  \ux\, invariant.

\section{Summary of results}
\hspace{0.8cm}\setcounter{equation}{0} We have shown that it is
possible to regulate supergravity at one loop by introducing
Pauli-Villars fields in chiral multiplets and Abelian gauge
multiplets.  For calculational simplicity, we restricted the dilaton
couplings to those of the classical limit of supergravity derived from
the heterotic string, but there is no impediment in principle to
extending our results to the more general case.  In the context of
string theory, this generalization is required, for example, when
nonperturbative string effects and/or GS terms are included in the
effective ``tree'' Lagrangian.  It would also be useful to know the
full one-loop corrections in the linear multiplet formulation.
However, certain one loop-effects such as the soft supersymmetry
breaking terms and the anomalous contributions to the Yang-Mills
kinetic term, depend only on gauged-charged matter and Yang-Mills
loops.  In this case, with the dilaton appearing only as a background
field, it is fairly straightforward~\cite{gnw,bgn} to include the
above-mentioned terms, and to generalize the results to the linear
multiplet formulation for the dilaton.  The A-terms for general 
supergravity without a GS term were calculated in Section 3, and were
found to be very sensitive to the details of the precise choice
of the Pauli-Villars couplings, which in turn can be determined only
with a detailed understanding of Planck-scale physics.

String-derived supergravity is anomalous at the quantum level under
perturbatively exact symmetries such as T-duality and \ux\ of the
underlying string theory.  When appropriate Green-Schwarz terms are
included, the effective field theory should be invariant, up to
nonperturbative string effects, at the quantum level. One could, for example,
make the regulated tree Lagrangian fully modular invariant by
including appropriate factors of the modular covariant Dedekind
function $\eta(iT)$ in the PV mass term $W_1$.  These would be
interpreted as threshold corrections from heavy string and
Kaluza-Klein modes. However, string-loop calculations show that at
least a part of the modular anomaly is canceled by a GS term; in
particular, for orbifolds like $Z_3$ and $Z_7$ with no $N=2$
supersymmetric twisted sector, there are no (modular noninvariant)
threshold corrections~\cite{ant} to the gauge kinetic term: $b^i_a =
0$ in (\ref{thr}). Moreover, cancellation of the \ux\, anomaly other
than by a GS mechanism seems problematic.

A part of the conformal anomaly can be directly inferred by replacing
$\ln \Lambda^2$ in (\ref{li}) by the real superfield $\ln
M^2(Z^i,\Z^{\m})$, where the lowest component $M^2(z^i,\z^{\m}) =
M^2(Z^i,\Z^{\m})|$ is the PV squared mass matrix.  Under a
transformation that leaves the regulated tree Lagrangian invariant
except for the PV mass terms, the shift in (\ref{li}) is determined by
[see (\ref{lprime})] $M^2(Z^i,\Z^{\m}) \to \tM^2(Z^i,\Z^{\m}) =
e^{H(Z) + \H(\Z)}M^2(Z^i,\Z^{\m})$, where $H(Z)$ is a holomorphic
function of the chiral fields.  The supersymmetric anomalies
associated with the F-term operators given in Section 2 are also
F-terms which contain the associated chiral anomalies; the general
form of these operators is given in Appendix 2. It has been
conjectured~\cite{anomalies} that all of these anomalies might be
canceled entirely or in part, depending on the string threshold
corrections in specific models, by the GS term included in
(\ref{gsterm}).  This would require a tree-level coupling of the
dilaton to the chiral superfields $\Phi_W,\Phi_\alpha$ in (\ref{ops}),
for example, inducing additional operators (and potential anomalies)
at the one-loop level.  The D-term operators of Section 2 give rise to
D-term anomalies, also displayed in Appendix A.  In principle these
could also be canceled by a tree-level coupling of the dilaton to real
superfields such as those in (\ref{tdt}) {\it via} a D-term of the
form (\ref{dterm}), again implying additional operators at one loop.
One such D-term is the shift in the K\"ahler potential,
(\ref{delK}). We have shown that the regularization of this term can
be made free of modular and \ux\ anomalies for supergravity from
orbifold compactification with background twisted sector fields set to
zero.  It is not clear that this can be achieved with twisted sector
fields in the background.

The full set of anomalous operators contains additional terms that
arise due to the fact that the PV masses are not constant; $D_\mu
\tM\ne0$.  Determining these requires keeping higher order terms in the
derivative expansion (as in the calculation of soft terms in Section
3.3) and retaining total derivatives (like the Gauss-Bonnet term) in
the coefficient of $\ln\Lambda^2$.  In addition it is necessary to
verify the cancellation
of linear divergences -- or equivalently\footnote{However this procedure
applied to modular and \ux\, anomalies will not insure, for example, the
correct dilaton dependence of the K\"ahler metric.} to show that 
(\ref{lprime}) is satisfied by comparing that expression with
with the anomaly calculated from $L(\Phi') - L(\Phi)$.  These issues
will be addressed elsewhere.  

\vskip .3in
\noindent{\bf Acknowledgements.} I wish thank Nima Arkani-Hamed,
Pierre Bin\'etruy, Joel Geidt, Hitoshi Murayama and Brent Nelson for
discussions, and to acknowledge the important contributions to this
work by my former collaborators Vidyut Jain and Kamran Saririan; this
paper is dedicated to Kamran's memory. This work was supported in part
by the Director, Office of Science, Office of High Energy and Nuclear
Physics, Division of High Energy Physics, of the U.S. Department of
Energy under Contract DE-AC03-76SF00098 and in part by the National
Science Foundation under grant PHY-95-14797.

\vskip .3in
\appendix

\def\ksubsection{\Alph{subsection}}
\def\theequation{\ksubsection.\arabic{equation}} 

     
\catcode`\@=11

\def\thesubsection{\Alph{subsection}.}
\def\thesubsubsection{\arabic{subsubsection}.}
\noindent{\large \bf Appendix}

\subsection{Component expressions of general superfields operators}
\setcounter{equation}{0}
After eliminating the auxiliary fields using their tree-level equations of motion~\cite{bggm}:
\beq F^i = - e^{-K/2}\A^i, \quad  \left.2R\right| = e^{-K/2}A, \quad -x{\bf D_a} = \D_a, \eeq
we obtain for the bosonic terms for the superfield operators introduced in Section 2.1:
\bea  \l\D_\beta T_\alpha\r &=& \epsilon_{\beta\alpha} T_0 + 
\(\sigma^{mn}\epsilon\)_{\beta\alpha}T_{mn}, \quad 
T_0 = {1\over2}\l\Dc T_\alpha\r, \quad T_{mn} = \epsilon_m^\mu\epsilon_n^\nu T_{\mu\nu},
\nonumber \\ \l\Dc T_\alpha\r &=& - 2D_{\m}T_i\(e^{-K}\A^iA^{\m} + 
\D_\mu z^i\D^\mu\z^{\m}\) + 2x^{-1}\D_aT_i(T^az)^i,\nonumber \\
T_{\mu\nu} &=& \[\(\D_\mu z^i\D_\nu\z^{\m} - 
\D_\nu z^i\D_\mu\z^{\m}\)D_{\m} - iF^a_{\mu\nu}(T_az)^i\]T_i,
\label{dt}\eea
where $m,n$ are tangent space Lorentz indices, and $\alpha,\beta$ are spinor indices
in the two-component spinor notation\footnote{The component field expressions 
use the metric $g_{\mu\nu} =$ diag$(+---)$, the opposite of the metric of~\cite{bggm}.}
of~\cite{bggm}.  For the bosonic
parts of the F-terms of Section 2.1 we obtain:
\bea L(W^\alpha_a T^a_\alpha) &=&  
 \[{\D_a\over x}\(\D_\rho z^i\D^\rho\z^{\m} + e^{-K}A^i\A^{\m}\) +
i\D_\mu z^i\D_\nu\z^{\m}F_{a-}^{\mu\nu}\]D_{\m}T^a_i \quad\nonumber \\ & & 
+  \cW_{ab}(T^bz)^iT^a_i + {\rm h.c.}, \label{tw} \eea\bea
& &L(T^\alpha T'_\alpha)\equiv L^{ij}T_iT'_j
= \(\cW^{ab} + \cbW^{ab}\)(T_az)^i(T_bz)^jT_iT'_j  \nonumber \\ & & \quad
+ \[{\D_a\over x}\(\D_\rho z^i\D^\rho\z^{\m} + e^{-K}A^i\A^{\m}\) +
i\D_\mu z^i\D_\nu\z^{\m}F_a^{\mu\nu}\](T^az)^j\(T_jD_{\m}T'_i + T_jD_{\m}T'_i\)
\nonumber \\ & & \quad - \(\D_\mu\z^{\m}\D^\mu z^i + e^{-K}A^{\m}\A^i\)
\(\D_\nu\z^{\n}\D^\nu z^j + e^{-K}A^{\n}\A^j\)D_{\m}T_iD_{\n}T'_j 
\nonumber \\ & & \quad
- \D^\mu z^i\D^\nu\z^{\m}\(\D_\mu z^j\D_\nu\z^{\n}
- \D_\mu\z^{\n}\D_\nu z^j\)D_{\m}T_iD_{\n}T'_j +{\rm h.c.}. \label{tt}\eea
In section 4 we also introduced F-terms of the form
\beq L(T, T')^\alpha_\alpha = L^{ij}T_jT'_i, \eeq
that is, they are they same as (\ref{tt}) except for the signs of 
two four-derivative terms.  In addition we have, with $X_{\mu\nu}= K_{\mu\nu}$
\bea L(6\Phi_W) &=& {1\over2}\int d^4\theta{E\over R} W^{\alpha\beta\gamma}
W_{\alpha\beta\gamma} + {\rm h.c.}\nonumber \\
&=&  {1\over2}\Da W_{\beta\gamma\delta}\D^\alpha W^{\beta\gamma\delta} 
+ {\rm h.c.} + {\rm fermions} \nonumber \\ 
&=& 6L_{GB} + {1\over4}r_{\mu\nu}r^{\mu\nu} - {1\over12}r^2
+ {1\over12}X_{\mu\nu}X^{\mu\nu} + {\rm fermions}.\eea
Up to terms that vanish on shell due to the graviton 
tree-level equations of motion, we have the identity [see (2.23)--(2.25) 
of~\cite{us}]
\bea {1\over12}\(3r_{\mu\nu}r^{\mu\nu} - r^2\) &=& 3L_\chi
- {1\over3}L_\alpha - {1\over12}X_{\mu\nu}X^{\mu\nu},\eea
and we obtain
\beq {1\over2}L_\chi + L_{GB} =
{1\over12}\int d^4\theta{E\over R}\(W^{\alpha\beta\gamma}
W_{\alpha\beta\gamma} - {1\over3}X_\alpha X^\alpha\) + {\rm h.c.}.\eeq
For the D-terms we obtain
\bea L(\phi_{WT}) &=& 
 \bigg(xF^{-a}_{\rho\mu}F_{+a}^{\rho\nu}\D_\nu z^i\D^\mu\z^{\m} 
+ 4e^{-K}\D\A^iA^{\m} \nonumber \\
& & + 2i\D_\mu z^i\D_\nu\z^{\m}\D^aF_a^{\mu\nu}  + 
2\D\D_\rho z^i\D^\rho\z^{\m} + \bigg)T_{i\m} ,\label{tim}\\ 
L(T^\alpha_\alpha) &=& 
 e^{-K}\(\A\D_\mu z^i\D^\mu\z^J + {1\over x}\D_a(T^az)^i\A^j\)t_{ij}
\nonumber \\ & &
+ e^{-K}\(e^{-K}A\A^i\A^j - {1\over2}\f^i\cbW\A^j\)t_{ij} \nonumber \\
& & - \(\D_\mu\z^{\m}\D^\mu z^i + e^{-K}A^{\m}\A^i\)D_{\m}(e^{-K}
\A^jt_{ij}) 
\nonumber \\ & & + e^{-K}\D_\mu\z^k\D^\mu z^i\A^j\(D_kt_{ij} - D_jt_{ik}\)
, \nonumber \\ T_{ij} &=& e^{-K/2}t_{ij}, \quad w(T_{ij}) = w(t_{ij})
- 2 = 2, \label{tij}\\ 
L(T^{\alpha\dot{\beta}}_{\alpha\dot{\beta}}) &=& 
\bigg(\D_\mu\z^{\m}\D_\nu z^i\D^\mu\z^{\n}\D^\nu z^j 
 + e^{-2K}A^{\m}\A^iA^{\n}\A^j\nonumber \\ & & + 2e^{-K}A^{\m}\A^i
\D_\nu\z^{\n}\D^\nu z^j\bigg)T_{ij\m\n},\label{tijmn}\eea
In the fully regulated Lagrangian, $\ln \Lambda^2$ in (\ref{li}) is
replaced by the real superfield $\ln M^2(Z^i,\Z^{\m})$, where
the lowest component $M^2(z^i,\z^{\m}) = M^2(Z^i,\Z^{\m})|$ is the PV
squared mass matrix.  Under a transformation that leaves the regulated 
tree Lagrangian invariant except for the PV mass terms:
$M^2(Z^i,\Z^{\m}) \to e^{H(Z) + \H(\Z)}M^2(Z^i,\Z^{\m})$, where
$H(Z)$ is a holomorphic function of the chiral fields,
the full anomaly associated with the one-loop generated F-term
operators given in Section 2 can be expressed in term of 
supersymmetric field operators of the form
\bea L(T,T',H) &=& 
{1\over2}\int d^4\theta {E\over R}T^\alpha T'_\alpha H(Z) 
+ {\rm h.c.}\nonumber \\
&=&  {1\over2}H(z)\Da T'_\beta\D^\alpha T^\beta + {\rm h.c.} + 
{\rm fermions} \nonumber \\ 
&=& - 2\re H T^0T'_0 - \re HT'_{\mu\nu}T^{\mu\nu} - \im H
{\tilde T}'_{\mu\nu}T^{\mu\nu} + {\rm fermions} \nonumber \\
&=& -2\re H L(T,T') - \im H {\tilde T}'_{\mu\nu}T^{\mu\nu} + 
{\rm fermions},\eea
where $L(T,T',1) 
= -{1\over2}L(T^\alpha T'_\alpha)$ is 
defined by (\ref{tt}), and 
\beq \re H\({1\over2}L_\chi + L_{GB}\) + \im H{r\tr\over48} =
{1\over12}\int d^4\theta{E\over R} H(Z)\(W^{\alpha\beta\gamma}
W_{\alpha\beta\gamma} - {1\over3}X_\alpha X^\alpha\) + {\rm h.c.}.\eeq
The chiral anomalies in the above expressions arise from the standard
nonlocal operators generated by fermion loops.
For the D-terms operators of Section 2, the corresponding anomalies
are also D-terms:
\beq L(\phi,H) = \int d^4\theta E\phi H + {\rm h.c.}.\eeq
In addition there are contributions from terms involving derivatives
of the Pauli-Villars masses that do not grow with the cut-off and were
not included in Section 2.

\subsection{Modifications of the $Z^I,Y_I$ contributions}
\setcounter{equation}{0}
The fields $\tZ^I_\alpha$ play the same role as $Z^I_1$ in I.
However, if we were to use the K\"ahler potentials $K^Z,K^Y$ adopted
in I, we would have for the covariant derivatives of the gauge kinetic
function $f(z) = s$: \beq f^{\tZ}_{IJ} = D_ID_Jf = - \Gamma^k_{IJ}f_k
\ne 0,\eeq which would generate unwanted contributions from
$\tZ_1^I$-loops.  The effect of the $f^i$-dependent terms in $K^{\tZ}$
is to eliminate these contributions; their presence in turn requires
compensating modifications of $K^{\tY}$ and $K^{\hY}$. In this
appendix we calculate the modifications with respect to I of the
$Z^I,Y_I$ loop contributions.

Denoting by a tilde quantities derived from the K\"ahler potentials
$K_\alpha^{Z,Y}$ in (\ref{kahl}) with $f_i=0,$ that is 
\beq \tK_\alpha^{Z,Y} = \l K_\alpha^{Z,Y}\r_{f_i = 0},\eeq
we have 
\bea L_3^Z + L_1^Y &=& \tL_3^Z + \tL_1^Y + \Delta\(L_3^Z + L_1^Y\), \eea
In I we found 
\bea \(L_3^Z + L_1^Y\)_I &=& - L_3 
- {2\over\sqrt{g}}e^{-K} \(A_i\A\L^i_I +  {\rm h.c.}\) \nonumber \\ & & - 
{2\over x\sqrt{g}}\[\D_a(T^az)^i\L_i^I + 
i\D_\mu\z^{\m}(T_az)^iK_{i\m}\L^{\mu a}_I + {\rm h.c.}\]\nonumber \\ & & 
+ 4\Delta_{\hV}L_I + 12\Delta_{M^2}L_I + 8\Delta_{\D}L_I ,\eea
where the subscript $I$ denotes the Lagrangian with $f=$ constant.  We have
\bea  A_i\L^i &=& A_i\L^i_I + x\cW A, \quad \Delta_{\hV}L = \Delta_{\hV}L_I ,
\quad \Delta_{M^2}L = \Delta_{M^2}L_I, \nonumber \\ 
{1\over x\sqrt{g}}\L^{\mu a} &=& {1\over x\sqrt{g}}\L^{\mu a}_I +
{\pp_\nu x\over x}F^{a\mu\nu} + {\pp_\nu y\over x}\tF^{a\mu\nu},\eea
and, from the results in (B.18) and (B.20) of~\cite{us2},
\bea && - {2\over x\sqrt{g}}\[\D_a(T^az)^i\L_i + {\rm h.c.}\] + 8\Delta_DL =
- {2\over x\sqrt{g}}\[\D_a(T^az)^i\L_i^I + {\rm h.c.}\]\nonumber \\ && \qquad 
+ 8\Delta_DL_I - 2{\pp_\mu x\over x}\D^aK_{j\m}\[\D^\mu z^j(T_a\z)^{\m} + 
(T_az)^j\D^\mu\z^{\m}\] + 32M^2\D \nonumber \\ && \qquad 
- 4i{\pp^\mu y\over x^2}\D^a\[K_{i\m}(T_az)^i\D_\mu\z^{\m} - {\rm h.c.}\]
+ {4\over x^2}\D\[\pp_\mu x\pp^\mu x + \pp_\mu y\pp^\mu y\] 
\nonumber \\ && \qquad + {\rm total\;derivative}. \eea
Combining these results, we obtain
\bea\tL_3^Z + \tL_1^Y &=& \(L_3^Z + L_1^Y\)_I + 2xM^2\(\cW + \cbW\)
- {1\over3}L(\Phi'_0),\label{mods}\eea
where we used (C.76) of~\cite{us2}.    

Writing
\bea K &=& k + G,\quad k = - \ln(s + \s),\quad k_i = - f_i/2x^2, \nonumber \\ 
K_{IJ} &=& \tK_{IJ} + \hK_{IJ} 
\quad \tK_{IJ} = K_{ij} - K_iK_j, \nonumber \\ 
\hK_{IJ} &=& - {1\over2x}\(f_iK_j + f_jK_i\) - {1\over2x^2}f_if_j,\eea
the effect of the $f_i$-dependent terms in $K^{\tZ}$ is to eliminate the
contributions to $K_{IJ}$ with $IJ = LS, SL,\; L\ne S$ 
(note that $\hK_{SS} = \tK_{SS} = K_{SS} = 0$).  
Since \beq 
\hK_{IJ}\tK^{IJ} = - \hK_{IJ}\hK^{IJ}, \eeq
we simply need to subtract the terms quadratic in $\hK_{IJ}$ in products of 
$K_{IJ}$ and its derivatives.
We have
\bea \hK_{IJ} &=& K_jk_i + K_jk_i - 2k_ik_j, \nonumber \\ 
\hGa^k_{IJ} &=& \delta^k_ik_j + \delta^k_jk_i + k^k_iK_j + k^k_jK_i 
- 4k^k_ik_j,\eea 
where $k^i_j = k^{i\m}k_{\m j}$ projects out $s$-components.  Then using 
\bea k_{i\m} &=& k_ik_{\m},\quad k_{ij} = k_ik_j,\quad D_jk_i = - k_ik_j, 
\quad k^iA_i = A, \quad k_i^jA_j = k_iA, \nonumber \\ 
A_s &=& k_sA,\quad A_{sk} = k_sA_k - k_sk_kA,
 \quad k^iA_{ij} = A_j - k_jA, \eea 
we obtain 
\bea \hA_{IJ} &=& \hK_{IJ}A - \hGa^k_{IJ}A_k
= - A_jk_i - A_ik_j + 2Ak_ik_j, \nonumber \\ 
\hR^k_{IJ\m} &=& \pp_{\m}\hGa^k_{IJ} =
\delta^k_ik_{j\m} + \delta^k_jk_{i\m} + k^k_iK_{j\m} +
k^k_jK_{i\m} - 4k^k_ik_{j\m}, \eea
and [with $\hat{k}_\alpha$, {\it etc.}, defined as in (\ref{tft}), (\ref{tdt})]
\bea &&R^{\dotb}_{\;\; IJk\dotb}R^{\alpha IJ}_{\;\;\;\;\;\;\alpha} =
\tR^{\dotb}_{\;\; IJk\dotb}\tR^{\alpha IJ}_{\;\;\;\;\;\;\alpha}
+ 8k^{\alpha\dotb}\(k_{\alpha\dotb} - K_{\alpha\dotb}\),\nonumber \\
&&\A_{IJ}R^{I\alpha J}_{\;\;\;\;\;\;\alpha} =
\btA_{IJ}\tR\A_{IJ}R^{I\alpha J}_{\;\;\;\;\;\;\alpha}
+ 4\A\hat{k}^\alpha\hat{k}_\alpha .\eea 
Then from the expression for $L_3$ given by Eqs. (\ref{dterm}),
(\ref{l3}), (\ref{tij}) and (\ref{tijmn}) [or
explicitly in (2.27) of I], we obtain, with $\eta^Z_1 = -1$,
\bea \Delta L_3^Z &=& -{\pp_\mu s\pp^\mu s\pp_\nu\s\pp^\nu\s\over2x^4} 
+ {2\over x^2}K_{i\m}\D_\mu z^i\D_\nu\z^{\m}\pp^\mu s\pp^\nu\s - 8M^4
\nonumber \\ & & - 12M^2\hV - 2M^2{\pp_\mu s\pp^\nu\s\over x^2} 
+ 2{e^{-K}\over x}\(\D_\mu z^i\pp^\mu s A_i\A + {\rm h.c.}\) \nonumber \\ & &
+ \({\pp^\mu s\pp_\mu s\over 2x^2} + {\rm h.c.}\)\(\hV + M^2\)
- 4e^{-K}\D_\mu z^i\D^\mu\z^{\m}A_i\A_{\m} \nonumber \\ & &
+ 2e^{-K}{\pp^\mu x\over x}\[\D_\mu z^i\(A_{ij}\A^j - 3A_i\A\) + {\rm h.c.}\]
- 4\hV^2 \nonumber \\ & & + 8M^2\D + 2e^{-K}\(\A A_{ij}\D_\mu z^j\D^\mu z^i - 
A_{ij}\A^j\A^iA + {\rm h.c.}\).\eea 

As noted in I, the derivatives of the metric defined by 
\beq K_{PQ}Y^P\Y^Q = \sum_{I,J=i,j}K^{i\bj} Y_I\Y_{\J} -
a\sum_{I=i}\(Y_I\Y^0\kappa^i + {\rm h.c.}\) 
+ |Y_0|^2\(1 + a^2\kappa^i\kappa_i\) ,\eeq
are most easily easily evaluated in terms of the derivatives of the inverse
metric
\beq K^{PQ}Y_P\Y_Q = \sum_{I,J=i,j}\(K_{i\bj} + a^2\kappa_i\kappa_{\bj}\)
\Y^IY^{\J} + a\sum_{I=i}\(\Y^IY^0\kappa_i + {\rm h.c.}\) + |Y_0|^2.\eeq
One finds for the elements of the affine connection
\bea \Gamma^I_{0k} &=& - aD_i\kappa_k + a^3K^{j\m}\kappa_j\kappa_i\pp_{\m}\kappa_k,
\quad \Gamma^0_{Ik} = -aK^{i\m}\pp_{\m}\kappa_k,\nonumber \\
\Gamma^I_{Jk} &=& - \Gamma^j_{ik} - a^2K^{j\m}\kappa_i\pp_{\m}\kappa_k,
\quad \Gamma^0_{0k} = a^2K^{i\m}\kappa_i\pp_{\m}\kappa_k.\eea
It follows immediately that $\Gamma^P_{P\alpha} = \tGa^P_{P\alpha}$,
so there are no changes to $H^2_Y,L^Y_2$. 
For $K^{\hY}$ we have $\kappa_i = K_i - k_i$, $(T_a)_Q^P\Gamma^Q_{P\alpha}
= (T_a)_Q^P\tGa^Q_{P\alpha}$, and
\bea D_I(T_ay)^J &=& \tD_I(T_ay)^J -
a^2k_j(T_az)^i, \quad D_I(T_ay)^0 = \tD_I(T_ay)^0, \nonumber \\
D_0(T_ay)^J &=& \tD_0(T_ay)^J - a^3k_j\D_a, \quad D_0(T_ay)^0 =
\tD_0(T_ay)^0, \nonumber \\ R^0_{0i\m} &=& \tR^0_{0i\m} -
a^2k_{i\m}, \quad R^0_{Ii\m} = \tR^0_{Ii\m} = 0, \nonumber \\
R^J_{Ik\m} &=& \tR^J_{Ik\m} + a^2\(\delta^i_kk_{j\m} + k^i_kK_{j\m} -
k^i_kk_{j\m}\), \eea 
with the result that for $P,Q = \hY$, 
\bea D_P(T_ay)^QD_Q(T_ay)^P &=& \tD_P(T_ay)^Q
\tD_Q(T_ay)^P, \nonumber \\ D_P(T_ay)^QR^P_{Qk\m} &=&
\tD_P(T_ay)^Q\tR^P_{Qk\m}, \eea
and the modifications to $L^{\hY}_1$ are determined by \bea
R^P_{Qk\m}R^Q_{Pj\n} &=& \tR^P_{Qk\m}\tR^Q_{Pj\n} + 2a^2 R^{(k)}_{k\m
j\n} \nonumber \\ & & + a^4\(2k_kk_{\m}k_jk_{\n} - k_{k\m}K_{j\n} -
K_{k\m}k_{j\n} - k_{j\m}K_{k\n} - K_{j\m}k_{k\n}\) \nonumber \\ &=&
\tR^P_{Qk\m}\tR^Q_{Pj\n} + 2\(a^4 - 2a^2\)k_kk_{\m}k_jk_{\n} \nonumber
\\ & & - a^4\(k_{k\m}K_{j\n} + K_{k\m}k_{j\n} + k_{j\m}K_{k\n} +
K_{j\m}k_{k\n}\),\label{riem}\eea 
where $R^{(k)}_{k\m j\n}$ is the Riemann tensor derived from $k$.

The K\"ahler metric for $\tY_S,\tY_0$ has $K^{i\m}\to k^{i\m}$ and $\kappa_i =
k_i$, and is the same as the $Y$-metric in I, with
the K\"ahler potential $K(z,\z)\to k(s,\s)$ and $a = 1$. Since
$(\Gamma^{\tY})^0_{S\alpha} = S_\alpha$, and 
$\Da S_\beta|$ has no bosonic terms
we need only consider
\beq (\Gamma^{\tY})^S_{S\alpha} = - \Gamma^s_{s\alpha} - a^2k_\alpha = 
- 3k_\alpha, \quad (\Gamma^{\tY})^0_{0 \alpha} = a^2 k_{\alpha}
= k_{\alpha},  \quad
\Phi^{\hY}_1 = 10\Phi_\beta.\label{riems} \eeq
Since $\sum_\alpha\eta^{\tY}_\alpha = -1$,
(\ref{riems}) gives a total contribution equal to $-10L_\beta$ to
$L^P_1$, (\ref{l21s}), but a portion $-4L_\beta$ of this is included in
$\beta'L_\beta$. Using (2.25) of I to evaluate the contribution from
(\ref{riem}), with $a^2\to a = -2,\;a^4\to a' = +2$, we obtain a net 
contribution: 
\bea \Delta L_1^Y &=& 3{\pp_\mu s\pp^\mu s\pp_\nu\s\pp^\nu\s\over4x^4} 
- {2\over x^2}K_{i\m}\D_\mu z^i\D_\nu\z^{\m}\pp^\mu s\pp^\nu\s 
 \nonumber \\ & & + \(3M^2 - \hV\){\pp_\mu s\pp^\mu\s\over x^2} 
+ 2{e^{-K}\over x}\(\D_\mu z^i\pp^\mu\s A_i\A + {\rm h.c.}\) \nonumber \\ & &
- 4M^2K_{i\m}\D_\mu z^i\D^\mu\z^{\m} - 4M^2\(3M^2 + 2\hV\)
- 6L_\beta,\eea 
and
\bea \Delta\(L_3^Z + L_1^Y\) &=& - 4M^2K_{i\m}\D_\mu z^i\D^\mu\z^{\m} 
+ {\pp_\mu s\pp^\mu s\pp_\nu\s\pp^\nu\s\over4x^4} 
- 4\hV^2 \nonumber \\ & &
- 20M^2\(M^2 + \hV\) + \(M^2 - \hV\){\pp_\mu s\pp^\nu\s\over x^2} 
\nonumber \\ & & + \({\pp^\mu s\pp_\mu s\over 2x^2} + {\rm h.c.}\)\(\hV + M^2\)
- 4e^{-K}\D_\mu z^i\D^\mu\z^{\m}A_i\A_{\m} \nonumber \\ & & 
+ 8M^2\D + 2e^{-K}\(\A A_{ij}\D_\mu z^j\D^\mu z^i - 
A_{ij}\A^j\A^iA + {\rm h.c.}\) \nonumber \\ & &
+ 2e^{-K}{\pp^\mu x\over x}\[\D_\mu z^i\(A_{ij}\A^j - A_i\A\) + {\rm h.c.}\]
- 6L_\beta.\eea
Using the relations [see (B.18) of \cite{us2}]
\bea {2\over\sqrt{g}}\L_i\A^iAe^{-K} +{\rm h.c.} &=& 
2e^{-2K}\(\D_\mu z^i\D^\mu z^j A_{ij}\A 
- A_{ij}\A^i\A^jA  + {\rm h.c.}\)
\nonumber \\ & & + 8M^2\(\hV + 3M^2 -\D\) + 2xM^2\(\cW + \cbW\) 
\nonumber \\ & & + 4e^{-K}\D^\mu z^i\D_\mu\z^{\m}(\A_iA_{\m} + K_{i\m}A\A) ,  
\nonumber \\ 2\pp_\mu\[\(\hV + M^2\){\pp^\mu x\over x}\] &=& 
2\pp_\mu\[e^{-K}{\pp^\mu x\over x}\(A_i\A^i - 2A\A\)\]\nonumber \\ &=&
2e^{-K}{\pp^\mu x\over x}\[\D_\mu z^i\(A_{ij}\A^j - A_i\A\) + {\rm h.c.}\]
\nonumber \\ & & + 2\(\hV + M^2\)\({\nabla^2 x\over x} - 
{\pp^\mu x\pp_\mu x\over x^2}\)\nonumber \\ &=& 
2e^{-K}{\pp^\mu x\over x}\[\D_\mu z^i\(A_{ij}\A^j - A_i\A\) + {\rm h.c.}\]
\nonumber \\ & & + \(\hV + M^2\)\({\pp_\mu s\pp^\mu s\over 2x^2} - 
{1\over x\sqrt{g}}f_i\L^i + x\cW + {\rm h.c.}\)\nonumber \\ & & 
+ 4\(\hV^2 + M^2\hV\) - \(\hV + M^2\){\pp_\mu s\pp^\mu\s\over x^2},
\nonumber \\ \Delta L_{M^2} &=& 
2e^{-K}{\pp^\mu x\over x}\(\D_\mu z^iA_i\A + {\rm h.c.}\) \nonumber \\ & &
- 2M^2\D + \hV^2 + 4M^2\hV + 6M^4,\eea
and dropping total derivatives we get
\bea \Delta\(L_3^Z + L_1^Y\) &=& {1\over\sqrt{g}}\lbr\L^i\[{f_i\over x}
\(\hV + M^2\) + 2e^{-K}A_i\A\] + {\rm h.c.}\rbr \nonumber \\ & & 
- 8\Delta L_{M^2} - 2L_\beta + 2x\hV\(\cW + \cbW\).\label{dels}\eea
Combining (\ref{dels}) with (\ref{mods}) gives the result in (\ref{l21s}),
with 
\bea -{1\over3}L(\Phi'_0) &=& - 32M^2\D - 
{4\over x^2}\D\pp_\mu s\pp^\mu\s 
- {2i\over x^2}\pp^\mu s\pp^\nu\s\D_aF^a_{\mu\nu} \nonumber \\ & &
+2\[\pp^\mu s\(iF^{-a}_{\mu\nu} + g_{\mu\nu}{1\over x}\D^a\)
K_{i\m}\D^\nu\z^{\m} + {\rm h.c.}\], \nonumber \\
L_\beta &=& M^4 + M^2{\pp_\mu s\pp_\nu\s\over2x^2}
+ {\pp_\mu s\pp^\mu s\pp_\nu\s\pp^\nu\s \over16x^4}.\label{kop}\eea

In addition we have, using $k^iW_i = 0$,
\bea A_{IJ}^{\tZ}\A^{IJ}_{\tZ} &=& \tA_{IJ}^Z\btA^{IJ}_Z - \A_{is}\A^{is} = 
A_{ij}\A^{ij} - 2\(A_i\A^i - A\A\) \nonumber \\
&=& A_{ij}\A^{ij} - 2e^K\(\hV + 2 M^2\),\nonumber \\
A^{{\hZ}_\alpha,\hY_\alpha}_{IJ} &=& A\delta^j_i, \quad 
\A_{{\hZ}_\alpha,\hY_\alpha}^{IJ} = \delta^i_j\A + a^2_\alpha
e^K\(K_j - k_j\)\(\A^i - k^i\A\), \nonumber \\ 
\A_{{\hZ}_\alpha,\hY_\alpha}^{I0} &=& a_\alpha \(\A^i - k^i\A\), \quad 
A^{{\hZ}_\alpha,\hY_\alpha}_{I0} = a_\alpha e^KW_i, 
\nonumber \\ A_{PQ}^{\hZ\hY}\A^{PQ}_{{\hZ}\hY} &=& 2\tA_{PQ}^{ZY}\btA^{PQ}_{ZY} 
- 2a^2_\alpha A\A, \quad A_{PQ}^{\tZ\tY}\A^{PQ}_{{\tZ}\tY} = 
2c^2_\alpha A\A ,\eea
giving the result in (\ref{pvkahl}). Note that the overall normalization of 
$A_{PQ}^{\hZ\hY}$ differs from that\footnote{There are extraneous factors of
$e^K$ and $W$ in the last term of the expression for
$\A_{Z_\alpha,Y_\alpha}^{IJ}$ in I.} used in I for $A^{ZY}_{PQ}$.   

Finally, there is a contribution from the diagonal part $f\delta_{ab}$ of the 
gauge kinetic function $f_{ab}$:
\bea f_{I0}^{\tZ\tY} &=& h_\alpha f_i, \quad \f^{I0}_{\tZ\tY} = h_\alpha\f^i, 
\quad \f^{IJ}_{\tZ\tY} = h_\alpha\f^ik_j, \nonumber \\
f_{PQ}^{\tZ\tY}\f^{PQ}_{\tZ\tY} &=& 2h^2_\alpha\f^if_i = 8x^2h^2_\alpha, \quad
\f^{PQ}_{\tZ\tY}A_{PQ}^{\tZ\tY} = 2h_\alpha c_\alpha\f^sk_sA = -4xA, 
\nonumber \\ A_{kIQ}^{\tZ\tY} &=& D_kA_{IQ}^{\tZ\tY} = \pp_kA_{IQ}^{\tZ\tY} - 
\Gamma^l_{ki}A_{LQ}^{\tZ\tY} - \(\Gamma^Y\)^P_{kQ}A_{IP}^{\tZ\tY} \nonumber \\
A_{kI0}^{\tZ\tY} &=& - a\(1 + a^2\)k_ik_kA = - 2k_ik_kA, \nonumber \\ 
A_{kIJ}^{\tZ\tY} &=& A_k + a^2k_k = \(A_k + k_kA\), \nonumber \\
\f^{PQ}_{\tZ\tY}A_{kPQ}^{\tZ\tY} &=& - 4xh_\alpha c_\alpha\(A_k - k_kA\).\eea
			
Then the scalar mass-matrix element $H^{\tZ\tY}_{PQ}$ takes the
form~\cite{us2} \bea H^{\tZ\tY}_{PQ} &=& e^{-K}\(A_{PQk}\A^k - A_{PQ}\A\)
+ {1\over2}f_{PQ}\cW, \nonumber \\ H^{\tZ\tY}_{PQ}H_{\tZ\tY}^{PQ} &=& -
2xh_\alpha c_\alpha \(\hV + M^2\)\(\cW + \cbW\) +
2x^2h^2_\alpha\cW\cbW\nonumber \\ & & + \cdots, \eea where the dots
represent contributions independent of $\cW$ that have already been
included.  Together with the results given in Appendix B of I, we obtain the 
contribution (\ref{lws}).

\subsection{Parameter constraints}
\setcounter{equation}{0} 

Defining
\beq \beta = \sum_C\eta^C\beta_C, \quad \sigma = 2\sum_C\eta^C\beta_C\alpha_C,
\eeq
if $\beta = - N_G -f$ and/or $\sigma\ne 0$ there are additional contributions
to the logarithmic divergences:
\bea \L_{PV} &\ni& \sqrt{g}\lln\[\(\sigma - {2\over3}\beta\)L'_\beta - \beta
L'_g\], \nonumber \\
L'_\beta &=& M^2\(\hV + 3M^2\)  + M^2 K_{i\m}\D_\mu z^i\D^\mu\z^{\m}
+ i{\pp_\mu s\pp_\nu\s\over4x^2}F^{\mu\nu}_a\D^a \nonumber \\ & &
+ 2\D M^2 + \(K_{i\m}\D_\mu z^i\D^\mu\z^{\m} + \hV + 3M^2 + 2\D\) 
{\pp_\nu s\pp^\nu\s\over4x^2} \nonumber \\ & &
+ {\pp_\mu s\pp_\nu\s\over4x^2}K_{i\m}\(\D^\mu z^i
\D^\nu\z^{\m} - \D^\mu\z^{\m}\D^\nu z^i\) ,\label{betal}\eea
where $L'_g$ is given in (\ref{pvls}).  The contribution from
(\ref{betal}) contains for example the terms
\bea \(\sigma - {2\over3}\beta\)L'_\beta - \beta L'_g &\ni& 
\(\sigma - 2\beta\)K_{i\m}\D_\mu z^i\D^\mu\z^{\m}{\pp_\nu s\pp^\nu\s\over4x^2}
\nonumber \\ & & - \(\sigma - {8\over3}\beta\)K_{i\m}\D^\mu\z^{\m}\D^\nu z^i
{\pp_\mu s\pp_\nu\s\over4x^2} ,\label{betall}\eea
that are not generated by any other contribution, requiring $\sigma = \beta=0$.

\subsection{The untwisted sector in orbifold compactifications}
\setcounter{equation}{0} Here we give explicitly the relations needed
for modular covariant ($K'_{PV} = K_{PV},\;W'_2 = e^{-F}W_2$)
regularization of the theory defined by (\ref{untwk}). Under the
modular transformation (\ref{fmod}) we have \bea K'_{p} &=&
N^{q}_{p}\(K_{q} + F_{q}\), \quad K'_{p\m} =
N^{q}_{p}N^{\n}_{\m}K_{q\n},\nonumber \\ K'_{PQ} &=&
N^{n}_{p}N^{m}_{q}\[K_{NM} - \sum_{i\ne j}\(F^i_{n}\Gj_{m} +
F^j_{m}\Gi_{n} + F^i_{m}F^j_{n}\)\], \nonumber \\ W'_{pq} &=&
e^{-F}N^{n}_{p}N^{m}_{q}\[W_{mn} - \sum_{i\ne j}\(F_{n}^jW_{m=ia} +
F_{m}^iW_{n=(aj)} - F^i_{m}F^j_{n}W\)\] \nonumber \\ W'_{p} &=&
e^{-F}N^{q}_{p}\(W_{q} - F_{q}W\).\eea 
The operators that determine
scalar curvature dependent quadratic divergences and the
logarithmically divergent contributions $L_{1,2}$ are: \bea
\Gamma_{p\alpha}^p &=& \tN G_\alpha, \quad
\Gamma_q^{p\alpha}\Gamma_{p\alpha}^q = \(\tN+1\)\Gic\Gia +
\hG^{p\alpha}_q\hG^q_{p\alpha}, \nonumber \\ \Gia &=&
-{1\over8}\(\DbDb - 8R\)\D_\alpha\Gi, \quad \hG^p_{q\alpha} =
-{1\over8}\(\DbDb - 8R\)\(\Gi_q\D_\alpha Z^p\) \delta_{i_p,i_q},
\nonumber \\ \tN &=& n + 2, \quad \Gamma_{p\alpha}^q(T^a)^p_q =
\sum_iT^a_i \Gia + G^a_\alpha, \quad T^a_i =
\sum_b(T^a)^{ib}_{ib}.\eea The
corresponding operators from $\tZ^{P},\;P\ne S$, are \bea &&
\(\Gamma_{\tZ}\)_Q^{P\alpha}\(\Gamma_{\tZ}\)_{P\alpha}^Q =
\Gamma_q^{p\alpha}\Gamma_{p\alpha}^q + \(2a^2 + a^4\) \(\Gic\Gia +
\hG^{p\alpha}_q\hG^q_{p\alpha}\) - 2\(a^2 +
a^4\)\hG_{p\alpha}Z^p_\alpha, \nonumber \\ && \hG_{p\alpha} =
-{1\over8}\(\DbDb - 8R\)G_p\Da \Gi\delta_{i_p,i}, \nonumber \\ 
&&\(\Gamma_{\tZ}\)_{P\alpha}^P =
\Gamma_{p\alpha}^p, \quad
\(\Gamma_{\tZ}\)_{P\alpha}^Q(T^a)^P_Q = \Gamma_{i\alpha}^j(T^a)^i_j +
a^2G^a_\alpha, \eea 
and the metric derivatives for $\hY_{P},\;P\ne S$, are related to
these by $\(\Gamma_{\hY}\)_{P\alpha}^Q = - \(\Gamma_{\tZ}\)_{Q\alpha}^P$.
The derivatives for $X' = \hZ,\tY$ are now related to
those for $X = \tZ,\hY$ by \bea
&&\(\Gamma_{X'}\)_Q^{P\alpha}\(\Gamma_{X'}\)_{P\alpha}^Q =
\(\Gamma_{X}\)_Q^{P\alpha}\(\Gamma_{X}\)_{P\alpha}^Q + \tN\(1 \pm
2\)G^\alpha G_\alpha, \nonumber \\ &&\(\Gamma_{X'}\)_{P\alpha}^P =
- \(\Gamma_{X}\)_{P\alpha}^P + \tN G_\alpha, 
\nonumber \\ &&\(\Gamma_{X'}\)_{P\alpha}^Q(T_a)^P_Q =
\(\Gamma_{X}\)_{P\alpha}^Q(T_a)^P_Q \pm X_\alpha\Tr T_a.\eea The
divergences from matter loops are canceled loops from $Z^I,Y_I$ and
$\phi^{(NI)},\; N = 0,P = 0,T,A=1,\cdots, n:$ \bea
\sum_I\Gamma_{(MI)}^{{(NI)}\alpha}\Gamma_{{(NI)}\alpha}^{(MI)} &=& \tN
G^\alpha G_\alpha , \quad \Gamma_{{(NI)}\alpha}^{(NI)} = \tN G_\alpha,
 \eea with additional
contributions that require a modification of the constraints on the
parameters $\alpha',\beta',\sigma$, as in Section 4.1, with $N\to3\tN$
in (\ref{newconst}). When an anomalous $U(1)$ is present we require
that some $\Phi_C$ carry $U(1)$ charge so as to cancel the last term in 
(D.4), as described in Section 5.2.

\subsection{Errata}
\setcounter{equation}{0} Here we list additional corrections
to~\cite{us2} that involve dilaton couplings, and were not reported in
I.
\begin{enumerate}
\item The second line of the RHS of the expression (C.48) for Tr$Y^2$ should 
read
$$ + {x^4\rho_i\rho^i\over8}\[\(F^a_{\mu\nu}F_b^{\mu\nu}\)^2
+ \(F^a_{\mu\nu}\tF_b^{\mu\nu}\)^2 - \(F^a_{\mu\nu}F_a^{\mu\nu}\)^2 -
\(F^a_{\mu\nu}\tF_a^{\mu\nu}\)^2\].$$
\item A contribution is missing from $T_3^{g+G}$ in (C.59), namely
$$ T_3^{g\alpha} = \[\hL_{\mu\nu},\m\]^\alpha_a\(M^{\mu\nu}\)^a_\alpha - 
\[\hL_{\mu\nu},m\]^\alpha_a\(\bM^{\mu\nu}\)^a_\alpha + 
(a\leftrightarrow \alpha) = {\pp_\mu x\pp_\nu y\over x^2}D^aF_a^{\mu\nu}.$$
\item There is a term missing from the expression (C.43), namely a contribtion
$$ - 3{\pp_\mu x\pp_\nu y\over x^2}\D^aF^a_{\mu\nu} $$
involving the graviton-gaugino connection in 
$2\(\tD_\mu\m\)^i_a\(\tD_\mu m\)_i^a$.
\item The sign of the first term on the RHS in the expression for 
$\tau_3^{\chi g}$ in (C.44) is incorrect.   
\item A contribution to $T^{g+G}$ is missing from (C.59), namely
$$ T^{g\alpha}_3 = {\pp_\mu x\pp_\nu y\over x^2}\D^aF^a_{\mu\nu} .$$
\item As noted in I, there are errors in the coefficients of $M^2\D$
in the traces given in Appendix C.  For the string dilaton
case considered here the changes with respect to the canoncial gauge kinetic 
energy case considered in I are: $-18$ in 
${1\over2}\STr H_\chi^2$, Eq. (C.36); $-14$ in 
${1\over8}\btr\(H_1^{\chi g}\)^2$, Eq. (C.41); $+ 2$ in $-T^{\chi g}_4$, Eq.
(C.44); $+58$ in ${1\over2}\STr H_{\chi g}^2$, Eq. (C.47); $+52$ in
${1\over2}\STr H_{\chi g}^2$, Eq. (C.62). \end{enumerate}

\end{document}